\documentclass[prd,preprintnumbers,nofootinbib]{revtex4} 
\usepackage{graphicx} 
\usepackage{amsmath}
\usepackage{amsfonts,amsbsy}
\usepackage{amssymb}
\usepackage{appendix}
\usepackage{slashed}

\def\lsim{ \,\, \vcenter{\hbox{$\buildrel{\displaystyle <}\over\sim$}}
 \,\,}
\def\be{\begin{equation}}
\def\ee{\end{equation}}
\def\bea{\begin{eqnarray}}
\def\eea{\end{eqnarray}}

\def\eq#1{{Eq.~(\ref{#1})}}
\def\fig#1{{Fig.~\ref{#1}}}
\def\fig#1{{Fig.~\ref{#1}}}
\def\s#1{{\slashed{#1}}}

\begin{document}

\title{\bf Diphoton production in high-energy p+A collisions}


\author{Alex Kovner$^{1}$ and Amir H. Rezaeian$^{2,3}$}
\affiliation{
$^1$ Dept. of Physics, University of Connecticut, High, Storrs, CT 06269, USA\\
$^2$ Departamento de F\'\i sica, Universidad T\'ecnica
Federico Santa Mar\'\i a, Avda. Espa\~na 1680,
Casilla 110-V, Valparaiso, Chile\\
$^3$  Centro Cient\'\i fico Tecnol\'ogico de Valpara\'\i so (CCTVal), Universidad T\'ecnica
Federico Santa Mar\'\i a, Casilla 110-V, Valpara\'\i so, Chile
}

\begin{abstract}

We consider semi-inclusive  diphoton+jet and inclusive diphoton production in high-energy proton-nucleus collisions, treating the target nucleus as a Color-Glass-Condensate and the projectile proton in the parton model. We obtain the prompt diphoton production cross-section in terms of fragmentation and direct contributions. The fragmentation part is given in terms of single-photon  and  double-photon fragmentation functions.  We study prompt, direct and fragmentation diphoton  correlations in p+p and p+A collisions at the LHC, and show that at low values of transverse momenta of the produced photon pair, these correlations are sensitive to saturation effects. We show that back-to-back  (de)-correlations  in prompt diphoton production are stronger in fragmentation part than in the direct one. 
\end{abstract}

\maketitle

\section{Introduction}
In recent years a lot of attention has been devoted to understanding of the physics of saturation \cite{sg}. Theoretical developments of the last 15-20 years have put this activity squarely on the first principle QCD foundations \cite{mv,jimwlk,bk,Balitsky:2008zza,Kovner:2013ona}. The saturation regime in hadronic scattering is qualitatively different from the simple parton model paradigm, and it would be extremely interesting to find clear signals of it in observed data. Its tell tale sign is the appearance of a dimensional saturation scale in a dense system of gluons (the Color Glass Condensate), which should dominate many bulk observables. 

Several pieces of experimental data at HERA \cite{jav1,amir-hera1,amir-hera2},  RHIC \cite{kn-rhic,dj-rhic,rhic-cgc}  and the LHC \cite{pp-LR,jav-d,pa-raju,jav-pa,pa-R,aa-LR,raju-glasma,hic-ap,pa-jam} have been indeed interpreted in the framework of saturation ideas, see also Refs.\,\cite{rhic-cgc1,rhic-cgc2,all-pa,jav-rev} and references therein.  However, one still cannot  point definitively to an  experimental verification of the saturation phenomenon. It is therefore important to understand what other observables can be sensitive to saturation \cite{na-sat}, and in particular to the existence of the saturation momentum. 

One aspect of saturation that has been discussed at length in recent literature, is its impact on particle correlations in the final state. This includes the effect on the ridge-like correlations  in rapidity and azimuthal angle \cite{ridge0,ridge1,ridge11,ridge2,ridge3,ridge33,ridge4,ridge5}, as well as decorrelation effect in forward dihadron \cite{dihadron}, photon-hadron \cite{amir-photon1,amir-photon2} and Drell-Yan lepton-pair-jet \cite{dy-ana} productions in high-energy proton-proton (p+p) and proton-nucleus (p+A) collisions. In this paper we continue the study of saturation effects on particle correlation.  

The main aim of this paper is to investigate diphoton azimuthal angular correlations at forward rapidity. The mechanism of diphoton production in the Color Glass Condensate (CGC) saturation framework is somewhat different than that of dihadron production. Soft gluons are scattered out of the projectile wave function by directly scattering on a saturated target, and via subsequent hadronization produce hadrons. Photons on the other hand do not scatter themselves, but rather decohere from the scattered quarks. It is thus interesting to see whether saturation has any discernible effects on the correlations between produced photons. 

In terms of theoretical description, there are clear advantages to studying prompt diphoton production as compared to dihadron production. For final state photons the difficulties involved with description of hadronization of final state quarks and gluons do not arise. For hadronic final states this stage of the process is usually described in terms of fragmentation functions, and this description is valid only at high transverse momentum. Additionally, one does not need to be concerned with possible initial state-final state interference effects which are present for hadron production. Within the CGC framework, the theoretical understanding of observables necessary to describe diphoton production is more robust. Unlike description of dihadron correlations, which necessitates the knowledge of correlators of a higher number of Wilson lines, the diphoton production cross section depends only on the dipole amplitude, which is the best understood observable in terms of high energy evolution.  

The diphoton production in proton-proton and antiproton-proton collisions has been intensively investigated in literature, see for example Refs.\,\cite{di-p0,di-p1, di-cut}. Precise theoretical understanding of the diphoton production in the standard model provides valuable guidance for the Higgs boson signal \cite{di-exp}. In the present paper, for the first time, we investiage diphoton production in high-energy proton-nucleus collisions. We obtain the prompt diphoton cross-section in the leading logarithmic approximation in terms of fragmentation and direct contributions, where the fragmentation part is given in terms of single-photon and  double-photon fragmentation functions.  We show that at low values of transverse momenta of the produced photon pair, back-to-back (de)-correlations in diphoton production are stronger in  fragmentation part than in the direct one. 

The plan of this paper is the following. In Sec. II we derive the basic formulae for calculating the cross-section of semi-inclusive diphoton+jet and inclusive diphoton production in high-energy proton-proton and proton-nucleus collisions in the CGC framework. The CGC approach is a first-principle effective field theory approach that describes the high-energy limit of QCD. In this formalism quantum corrections enhanced by large logarithms of $1/x$ are systematically re-summed incorporating high gluon density effects at small x and for large nuclei \cite{mv,jimwlk,bk}. In Sec. II, we also discuss the soft limit, in which the expressions simplify and become amenable to numerical calculations. As an illustrative example, in Sec. II, we also obtain the cross-section of single inclusive prompt photon production in the soft approximation.  In Sec. III we present the results of numerical calculations for correlations in direct, fragmentation and prompt diphoton production, together with a short discussion. We summarize our main results in Sec. IV.

\section{Semi-inclusive diphoton+jet production in proton-nucleus collisions}
\begin{figure}[t]                                       
                                  \includegraphics[width=8 cm] {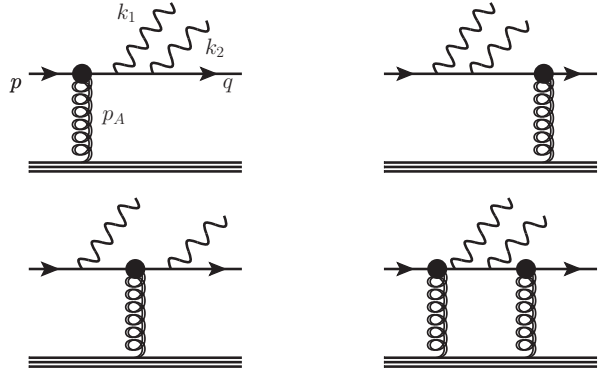} 
\caption{The diagrams contributing to diphoton production of a quark in the background of the CGC field. The black blob denotes the interaction of a quark to all orders with the background field via multiple gluon exchanges.  See Appendix for the definition of kinematics.  }
\label{f1}
\end{figure}

In this section we introduce the basic formulae for calculating the cross-section for the following process: $h+A\to \gamma_1\,+\,\gamma_2\,+\, X$,  where a dilute projectile hadron interacts coherently with a dense target $A$ and produces two photons $\gamma_1$ and $\gamma_2$. In the leading order approximation, at forward rapidity,  a valence quark of the projectile hadron emits two photons via Bremsstrahlung and the produced diphoton+jet is then put on shell by 
interacting coherently over the whole longitudinal extent of the target. The cross section for production of a quark  with momentum $q$ and two prompt photons with momenta $k_1$ and $k_2$  in the scattering of an on-shell quark with momentum $p$ off a hadronic target (either a proton or a nucleus),
\be
q(p)+A\to \gamma(k_1) +\gamma(k_2) + \text{jet}(q) + X,
\ee
can be written in the following general form, 
\be d\, \sigma^{q\to q\gamma \gamma} = \frac{d^3 k_1}{(2\pi)^3\,2 k_1^-}  \frac{d^3 k_2}{(2\pi)^3\,2 k_2^-} \frac{d^3 q}{(2\pi)^3\, 2
  q^-} \frac{1}{2 p^-}(2\pi)\, |\mathcal{M}(p|q,k_1,k_2)|^2\delta (p^- - q^- -k_1^- -k_2^-), 
\label{m1}
\ee
where the matrix element $ \mathcal{M}$ is related to the scattering amplitude by 
\be 
\langle q(q),\gamma(k_1),\gamma(k_2)|q(p)\rangle=(2\pi)\,\delta (p^- - q^- -k_1^- -k_2^-)\mathcal{M}(p|q,k_1,k_2).  
\label{m2}
\ee
In the CGC approach, we assume that the small-x gluon modes of the nucleus have a large occupation number so that it can be described in terms of a classical color field.  This should be a good approximation for large enough nucleus at high-energy\footnote{Note that proton at high-energy and specially at very forward rapidity is a dense system and in principle the same approximation also applies there, see for example Refs.~\cite{jav1,amir-hera1,amir-hera2,kn-rhic,dj-rhic,rhic-cgc,pp-LR,jav-d,pa-raju,jav-pa,pa-R,aa-LR,raju-glasma,hic-ap,pa-jam}.}. This color field emerges from the classical Yang-Mills equation with a source term provided by faster partons. The renormalization group equations which govern the separation between the soft and hard models are then given by the non-linear Jalilian-Marian, Iancu, McLerran, Weigert, Leonidov, Kovner (JIMWLK) evolution equations \cite{jimwlk} (see below).  We further assume that the projectile proton is in the dilute regime and can be described in ordinary perturbative approach, in terms of parton distribution functions. In this framework, the scattering amplitude of diphoton+jet production in quark-nucleus scatterings in momentum space in lowest order in the electro-magnetic $\alpha_{em}$ and the strong $\alpha_s$ coupling constants can be written in the following formal form, 
\bea
\langle q({\bf q}),\gamma({\bf k_1}),\gamma({\bf k_2})|q({\bf p})\rangle
&=&-ie_q^2\bar{u}({\bf q})\Big[\mathcal{T}_F(q;p-k_1-k_2)G_F^0(p-k_1-k_2)\slashed{\epsilon}(k_2)G_F^0(p-k_1)\slashed{\epsilon}(k_1)\nonumber\\
&+&  \slashed{\epsilon}(k_2)G_F^0(q+k_2)\slashed{\epsilon}(k_1)G_F^0(q+k_1+k_2)\mathcal{T}_F(q+k_1+k_2;p)\nonumber\\
&+&  \slashed{\epsilon}(k_2)G_F^0(q+k_2)\mathcal{T}_F(q+k_2;p-k_1) G_F^0(p-k_1)\slashed{\epsilon}(k_1)\nonumber\\
&+&  (k_{1}\leftrightarrow k_{2}) \Big]u({\bf p}), \label{am1} \ 
\eea
where $e_q$ is the fractional electric charge of the projectile quark, $G_F^0$ is the free Feynman propagator of a quark, $u$ and $\epsilon_\mu$ denote the quark free spinor and the photon polarization vector respectively. In the above, the operator matrix $\mathcal{T}_F$ contains the interaction between the quark and the colored glass condensate target, which resums multiple interactions with the background CGC field \cite{g-cgc0,g-cgc1}.  Assuming that the target is moving in the positive $z$ direction, we have \cite{pho-cgc1}, 
\be 
\mathcal{T}_F(q;p)=2\pi\delta(q^--p^-)\gamma^- sign(p^-)\int d^2{\bf z_T}\big[U({\bf z_T})-1\big]e^{i({\bf q_T-p_T})\cdot{\bf z_T}}, \label{tf}
\ee 
where $U(z_T)$ is a unitary matrix in fundamental representation of $SU(N_c)$ - the scattering matrix of a quark on the colored glass condensate target:
\be 
U({\bf z_T})=T \exp\left(-ig^2\int dx^-\frac{1}{\nabla^2_T}\rho_a(x^-,{\bf x_T})t^a\right).
\ee 
Here $\rho$ is the density of the color sources in the target and $t^a$ is the generator of $SU(N_c)$ in the fundamental representation. The expression in \eq{am1}  accounts for all processes illustrated in \fig{f1} for diphoton+jet production of a quark interacting with the CGC background field at lowest order in $\alpha_{em}$ and $\alpha_s$ in the  leading-log approximation. Note that the lower diagram on the right panel in \fig{f1} does not contribute at high energy. 
Conceptually this is because the nucleus is moving with speed of light in the $+z$ direction. 
By the time the diphoton is emitted from the quark, the nucleus has already moved far away from the quark and no further interactions are allowed by causality.

Using the definition of $\mathcal{T}_F$ in \eq{tf}, one can rewrite the amplitude as, 
\bea
\langle q({\bf q}),\gamma({\bf k_1}),\gamma({\bf k_2})|q({\bf p})\rangle &=&ie_q^2\bar{u}({\bf q})\Big[\frac{\gamma^-
   (\s{p}-\s{k_1}-\s{k_2})\s{\epsilon}(k_2)(\s{p}-\s{k_1})\s{\epsilon}(k_1)}
{(p-k_1-k_2)^2 (p-k_1)^2}
+\frac{\s{\epsilon}(k_2)
   (\s{q}+\s{k_2})\s{\epsilon}(k_1)(\s{q}+\s{k_1}+\s{k_2}) \gamma^-}
{(q+k_2)^2(q+k_1+k_2)^2}\nonumber\\
&+&\frac{\s{\epsilon}(k_2)(\s{q}+\s{k_2})\gamma^-(\s{p}-\s{k_1})\s{\epsilon}(k_1)}
{(q+k_2)^2(p-k_1)^2} + (k_{1}\leftrightarrow k_{2})\Big]u({\bf p})\nonumber\\
&\times& 2\pi \delta(q^-+k^-_1+k^-_2-p^-)\int d^2{\bf z_T}\big[U({\bf z_T})-1\big]e^{i({\bf q_T+k_{1T}+k_{2T}-p_T})\cdot{\bf z_T}}. \label{am-di}
\eea
The  semi-inclusive diphoton+jet cross-section defined in \eq{m1} can be readily obtained by squaring the amplitude and averaging over the color charge distribution. To this end, one needs to perform the color charge averaging of the expersion $\langle U^{\dag}({\bf x_T})U({\bf y_T})\rangle_{\rho}$ with the CGC weight
\be 
W[\rho]=T \exp\left(-\int dx^-d^2{\bf x_T}\frac{\rho_a(x^-,{\bf x_T})\rho^a(x^-,{\bf x_T})}{2\mu^2(x^-)}    \right).
\ee 
Note that the averaging procedure does not affect the spin dependence in \eq{m1}. Therefore, one can  rewrite the final expression in term of the dipole-target forward scattering amplitude $N_F$ and a spin trace, 
\bea d\, \sigma^{q\to q\gamma \gamma} &=& \frac{e_q^4}{2}\frac{d^3 {\bf k_1}}{(2\pi)^3\,2 k_1^-}  \frac{d^3 {\bf k_2}}{(2\pi)^3\,2 k_2^-} \frac{d^3{\bf  q}}{(2\pi)^3\, 2
  q^-} \frac{1}{2 p^-}(2\pi)\, \delta (p^- - q^- -k_1^- -k_2^-) \langle \text{tr}(S^{\dag} S)\rangle_{spin}\nonumber\\
&\times& d^2{\bf r_T}d^2{\bf b_T}  e^{i({\bf p_T}-{\bf q_T}-{\bf k_{1T}}-{\bf k_{2T}})\cdot{\bf r_T}} N_F({\bf b_T}, {\bf r_T},x_ g), 
\label{m22}
\eea
where the factor $1/2$ is due to averaging over flavor SU(2) and $N_F$ is the imaginary part of (quark-antiquark) dipole-target forward scattering amplitude defined as 
\be
N_F({\bf b_T},{\bf r_T},x_g) = \frac{1}{N_c} \, < Tr [1 - U^{\dagger} ({\bf x_T}) U ({\bf y_T}) ] >.
\label{cs_def}
\ee
Here $N_c$ is the number of colors, the vector ${\bf b_T}\equiv ({\bf x_T} + {\bf y_T})/2$ is the impact parameter of the dipole relative to the target and ${\bf r_T}\equiv {\bf x_T} - {\bf y_T}$ is the dipole transverse vector. The dependence of the dipole scattering probability  on Bjorken $x_g$ is determined by the JIMWLK renormalization group equations (see Sec. III).  The explicit expression for the trace in \eq{m2} is, 

\bea
 \langle \text{tr}(S^{\dag} S)\rangle_{spin}  &=&\frac{1}{2}
\text{tr} \Bigg\{\s{p}\Bigg[\frac{\s{\epsilon}^*(k_1) (\s{p}-\s{k_1}) 
  \s{\epsilon}^*(k_2)  (\s{p}-\s{k_1}-\s{k_2}) \gamma^-}
{(p-k_1-k_2)^2 (p-k_1)^2}
+\frac{\gamma^-
  (\s{q}+\s{k_1}+\s{k_2})  \s{\epsilon}^*(k_1)  (\s{q}+\s{k_2})\s{\epsilon}^*(k_2)}
{(q+k_2)^2(q+k_1+k_2)^2}\nonumber\\
&+&\frac{\s{\epsilon}^*(k_1) (\s{p}-\s{k_1}) \gamma^- (\s{q}+\s{k_2})\s{\epsilon}^*(k_2)}
{(q+k_1)^2(p-k_2)^2} +  \frac{\s{\epsilon}^*(k_2) 
  (\s{p}-\s{k_2})\s{\epsilon}^*(k_1)  (\s{p}-\s{k_1}-\s{k_2}) \gamma^-}
{(p-k_1-k_2)^2 (p-k_2)^2}    \nonumber\\
&+&\frac{\gamma^-
(\s{q}+\s{k_1}+\s{k_2})  \s{\epsilon}^*(k_2)  (\s{q}+\s{k_1}) \s{\epsilon}^*(k_1)}
{(q+k_1)^2(q+k_1+k_2)^2}
+\frac{\s{\epsilon}^*(k_2)(\s{p}-\s{k_2}) \gamma^- (\s{q}+\s{k_1})\s{\epsilon}^*(k_1)}
{(q+k_2)^2(p-k_1)^2} \Bigg] \nonumber\\
&\times&
\s{q}\Bigg[\frac{\gamma^-
   (\s{p}-\s{k_1}-\s{k_2})\s{\epsilon}(k_2)(\s{p}-\s{k_1})\s{\epsilon}(k_1)}
{(p-k_1-k_2)^2 (p-k_1)^2}
+\frac{\s{\epsilon}(k_2)
   (\s{q}+\s{k_2})\s{\epsilon}(k_1)(\s{q}+\s{k_1}+\s{k_2}) \gamma^-}
{(q+k_2)^2(q+k_1+k_2)^2}\nonumber\\
&+&\frac{\s{\epsilon}(k_2)(\s{q}+\s{k_2})\gamma^-(\s{p}-\s{k_1})\s{\epsilon}(k_1)}
{(q+k_2)^2(p-k_1)^2} +
\frac{\gamma^- (\s{p}-\s{k_1}-\s{k_2})\s{\epsilon}(k_1)(\s{p}-\s{k_2})\s{\epsilon}(k_2)}
{(p-k_1-k_2)^2 (p-k_2)^2}\nonumber\\
&+&\frac{\s{\epsilon}(k_1)
   (\s{q}+\s{k_1})\s{\epsilon}(k_2)(\s{q}+\s{k_1}+\s{k_2}) \gamma^-}
{(q+k_1)^2(q+k_1+k_2)^2}
+\frac{\s{\epsilon}(k_1)(\s{q}+\s{k_1})\gamma^-(\s{p}-\s{k_2})\s{\epsilon}(k_2)}
{(q+k_1)^2(p-k_2)^2}
\Bigg]    \Bigg\}, \label{tt}\
\eea
where the factor $1/2$ is due to averaging over the spin of the projectile quark. 
The trace in \eq{tt} can be somewhat simplified by summing over the photon polarization and using the relation $\epsilon_\mu({\bf k})\epsilon_\nu^*({\bf k})=-g_{\mu\nu}$  (note that terms proportional to $k_\mu$  do not contribute due to Ward identities). Moreover, half of the terms in \eq{tt} are symmetric under the replacement of $k_1\to k_2, k_2\to k_1$.  Nevertheless, even after these simplifications, the exact expression for the trace in \eq{tt} is rather complicated and  very difficult for a numerical evaluation. However, one can simplify it significantly by restricting to the soft limit, which is  relevant for the high-energy collisions (see below).

\subsection{Single-inclusive prompt photon production: the soft limit}
The calculation of the photon+jet and diphoton+jet production in the CGC approach in the soft limit is rather similar. Therefore, it is instructive to first derive the cross-section of semi-inclusive photon+jet production. 

Let us consider production of a single prompt photon and a quark with 4-momenta $k$ and $q$ respectively in scattering of a on-shell quark with 4-momentum $p$ on a nuclear (or proton) target in the CGC approach in the soft approximation namely when $|k|<|p-q|$.  To this end, one can calculate the amplitude from  diagrams similar to those shown in the upper panel of \fig{f1}  replacing two photon lines by a single one with momentum $k$ \cite{pho-cgc1}:  
\bea
\langle q({\bf q}),\gamma({\bf k})|q({\bf p})\rangle&=&-i e_q\bar{u}({\bf q})\Big[\frac{\gamma^-(\s{p}-\s{k})\s{\epsilon}}{(p-k)^2}+ \frac{\s{\epsilon}(\s{q}+\s{k})\gamma^-}{(q+k)^2} \Big]u({\bf p})\, 2\pi \delta(q^-+k^- -p^-) \nonumber \\
&\times&\int d^2{\bf z}_T\big[U({\bf z}_T)-1\big]e^{i({\bf q}_T+{\bf k}_{T}-{\bf p}_T)\cdot{\bf z}_T},  \nonumber \\
&\approx& -ie_q\bar{u}({\bf q})\gamma^- u({\bf p})\Big[\frac{q\cdot\epsilon}{q\cdot k}-\frac{p\cdot\epsilon}{p\cdot k}\Big]\, 2\pi \delta(q^-+k^- -p^-)\int d^2{\bf z}_T\big[U({\bf z}_T)-1\big]e^{i({\bf q}_T+{\bf k}_{T}-{\bf p}_T)\cdot{\bf z}_T}.\
\eea
In the above equation second line, we implemented the soft limit approximation and used
\begin{equation}\label{soft-a}
\s{p} \s{\epsilon} u({\bf p}) = 2 p\cdot\epsilon\, u({\bf p}),  \hspace{2cm} \bar{u}({\bf q})  \s{\epsilon} \s{q} =\bar{u}({\bf q})\, 2 q\cdot\epsilon, \hspace{2cm} 
(p-k)^2\approx -2p\cdot k.  \
\end{equation}
 The spinor averaged matrix element can be then immediately obtained, 
\bea \label{soft-d}
 \langle \text{tr}(S^{\dag} S)\rangle_{spin}^{\text{single-photon, soft}}  &=&\frac{1}{2}
\text{tr}\{\s{q}\gamma^-\s{p}\gamma^-\}
\Big{|}\frac{q\cdot \epsilon(k)}{q\cdot k}-\frac{p\cdot \epsilon(k)}{p\cdot k}\Big{|}^2=16p^-q^-\Big[\frac{(p\cdot q)}{(q \cdot k)(p\cdot k)}\Big],\\
&=& \frac{32\,p^- q^- k^{-2} {\bf q}^2_T}{{\bf k}_T^2(k^-{\bf q}_T-q^-{\bf k}_T)^2}.\label{spin-single}\
\eea
In Eq.\,(\ref{soft-d}) summations over the spin of the final quark and over the polarization of the photon were performed.  
The cross-section of single inclusive prompt photon production, similar to \eq{m22}, can be written as 

\bea d\, \sigma^{q\to q \gamma} &=& \frac{e_q^2}{2} \frac{d^3 {\bf k}}{(2\pi)^3\,2 k^-} \frac{d^3{\bf  q}}{(2\pi)^3\, 2
  q^-} \frac{1}{2 p^-}(2\pi)\, \delta (p^- - q^--k^-) \langle \text{tr}(S^{\dag} S)\rangle_{spin}^{\text{single-photon}}\nonumber\\
&\times& d^2{\bf r}_Td^2{\bf b}_T  e^{i({\bf p}_T-{\bf q}_T-{\bf k}_{T}).{\bf r}_T} N_F({\bf b}_T, {\bf r}_T,x_ g). 
\label{m22-single}
\eea
Using \eq{spin-single} and the definitions of the rapidities of produced photon $\eta_\gamma=\log\left (\frac{\sqrt{2}k^-}{k_T}\right)$ and quark $\eta_h=\log\left (\frac{\sqrt{2}q^-}{q_T}\right)$, the above equation can be simplified in the soft limit to yield
\bea  \label{p-h-1}
\frac{d\sigma^{q(p)\to q(q) \gamma(k) X}}{dk^2_T d\eta_{\gamma}dq_T^2 d\eta_h d\theta}&=&\frac{2\alpha_{em}e^2_q}{(2\pi)^3\sqrt{2s}} \frac{q^- k^{-2} q^2_T}{ k_T^2(k^-{\bf q}_T-q^-{\bf k}_T)^2}\delta (x_q-\frac{k_T}{\sqrt{s}} e^{\eta_{\gamma}}-\frac{q_T}{\sqrt{s}} e^{\eta_{h}} ) \times\nonumber\\
&&\int d^2{\bf r}_Td^2{\bf b}_T  e^{i({\bf q}_T+{\bf k}_T).{\bf r}_T} N_F({\bf b}_T, {\bf r}_T,x_ g),\
\eea
where $\theta$ is the angle between the produced jet and photon. The parameter $x_q$ is the ratio of energies of the incoming quark to nucleon,  $x_q=p^-/\sqrt{s/2}$ with $\sqrt{s}$ being the nucleon-nucleon center-of-mass energy. 

In order to relate the above partonic production cross-section to the cross section of photon-hadron production in 
proton-nucleus collisions, one needs to convolute the 
partonic cross-section with the quark and
antiquark distribution functions of a proton and the quark-hadron 
fragmentation function, 
\begin{eqnarray}\label{qh-f}
\frac{d\sigma^{q(p)\to h(q^\prime) \gamma(k)}}{d^2{\bf k}_T d\eta_{\gamma}d^2{\bf q}^\prime_T d\eta_h d\theta}&=& \int^1_{z_{h}^{min}} \frac{dz_h}{z_h^2} \, 
 \int\, dx_q\, f (x_q,\mu_I^2)  \frac{d\sigma^{q(p)\to q(q) \gamma(k)}}{d^2{\bf k}_T d\eta_{\gamma}d^2{\bf q}_T d\eta_h d\theta}
  D_{h/q}(z_h,\mu_F^2),
\end{eqnarray}
where $q_T^\prime$ is the transverse momentum of the produced hadron, and $f(x_q,\mu_I^2)$ is the 
parton distribution function (PDF) of the incoming proton which depends on the light-cone momentum fraction $x_q$ and the hard-scale $\mu_I$. Summation over the 
quark and antiquark flavors in the above expression is understood.  The function $D_{h/q}(z_f,\mu_F^2)$ is the quark-hadron fragmentation function (FF) 
where $z_h$ is the ratio of energies of the produced hadron and quark and $\mu_F$ is the fragmentation scale. The produced hadrons are assumed here to be massless.
The light-cone momentum fractions $x_q, x_g, z_h$ are related to the transverse momenta and
rapidities of the produced hadron and prompt photon via, 
\begin{eqnarray}\label{qh-k}
x_q&=&x_{\bar{q}}=\frac{1}{\sqrt{s}}\left( k_T\, e^{\eta_{\gamma}}+\frac{q^\prime_T}{z_h}\, e^{\eta_{h}}\right),\nonumber\\
x_g&=&\frac{1}{\sqrt{s}}\left(k_T\, e^{-\eta_{\gamma}}+ \frac{q^\prime_T}{z_h}\, e^{-\eta_{h}}\right),\nonumber\\
z_h&=&q_T^\prime/q_T \hspace{1 cm} \text{with}~~~~~ z_{h}^{min}=\frac{q_T^\prime}{\sqrt{s}}
\left(\frac{e^{\eta_h}}
{1 - \frac{k_T}{\sqrt{s}}\, e^{\eta_{\gamma}}}\, 
\right).\
\end{eqnarray}
In order to obtain the cross-section for the single inclusive prompt photon production, we integrate over the outgoing quark momentum in \eq{p-h-1}. Using $d\eta_h=dq^-/q^-$ we obtain,
\begin{equation} \label{s-p}
\frac{d\sigma^{q(p)\to \gamma(k) X}}{d^2{\bf k}_T d\eta_{\gamma}}=\frac{2\alpha_{em}e^2_q}{(2\pi)^3k_T^2} \int d^2 {\bf q}_T \frac{k^{-2} {\bf q}^2_T}{(k^-{\bf q}_T-q^-{\bf k}_T)^2}
N_F(|{\bf q}_T+{\bf k}_T|, x_g).
\end{equation}
In terms of the photon fragmentation parameter $z=k^-/p^-$, the light-cone fraction variable $x_g$ in \eq{s-p} is then expressed as, 
\begin{equation}
 x_g=\frac{1}{x_q s}\big[\frac{k_T^2}{z}+\frac{ q_T^2}{1-z}\Big]. \label{xg-s}
\end{equation}
The collinear singular part in \eq{s-p}  is naturally attributed to the fragmentation contribution. Shifting the momentum ${\bf q}_T\to {\bf q}_T+ {\bf k}_T/z$ and  breaking the integral into two parts by introducing a hard cutoff, the cross-section of single inclusive prompt photon can be written as a sum of the fragmentation and the  direct photon (finite)  part: 
\begin{equation} \label{s-p-2}
\frac{d\sigma^{q(p)\to \gamma(k) X}}{d^2{\bf k}_T d\eta_{\gamma}}= \frac{\alpha_{em}e^2_q}{\pi (2\pi)^3} 
\frac{2}{ k_T^2}\int_{q_T^2>\mu_F^2} d^2 {\bf q}_T\, \frac{|{\bf q}_T+{\bf k}_T/z|^2}{q_T^2}\, N_F(|{\bf q}_T+{\bf k}_T(1+1/z)|, x_{1g})+\frac{1}{(2\pi)^2}\frac{1}{z}D_{\gamma/h}(z,\mu_F^2) N_F(k_T/z,x_{2g}),
\end{equation}
where the fragmentation scale $\mu_F$ used to separate the soft from hard contribution. 
The first term is the direct photon contribution, whereas the second term is the fragmentation photon contribution, corresponding to the kinematics where the photon is emitted almost collinearly with the outgoing quark. The photon fragmentation function extracted from \eq{s-p} in the soft approximation is given by
\begin{equation} \label{s-ff}
D_{\gamma/h}(z,\mu_F^2)=\frac{\alpha_{em}e^2_q}{2\pi}\frac{2}{z}\log\left(\mu_F^2/\Lambda^2_{QCD}\right).
\end{equation}
The light-cone fraction variables $x_{1g}$ and  $x_{2g}$ in \eq{s-p-2} are obtained via \eq{xg-s} by replacing ${\bf q}_T\to {\bf q}_T+ {\bf k}_T/z$  and ${\bf q}_T\to {\bf k}_T/z$, respectively. 
The above expression for the single-inclusive photon fragmentation function agrees with the corresponding expression obtained in the standard perturbative QCD calculation in the leading-log approximation in the soft limit \cite{own,amir-photon1}. Note that in the soft photon approximation we assumed $z<<1$ and $k_T<<q_T$.  The cross-section given in \eq{s-p-2} is also in accordance with expression obtained in Refs.\,\cite{pho-cgc1,pho-cgc2,amir-photon1} in the soft limit. 

Note that both direct and fragmentation cross-section for single inclusive photon production are explicitly proportional to the dipole amplitude. Thus in principle they probe the small-x dynamics and saturation physics in the appropriate kinematics \cite{amir-photon1,amir-photon2} (see also Ref.\,\cite{pho-all}).

\subsection{Semi-inclusive diphoton+jet production: soft limit }
We now turn to the problem of diphoton+jet production in proton-nucleus collisions assuming that the radiated photons are soft, namely $|k_{1,2}|<|p-q|$. In this case we can simplify the expression in \eq{am-di} by ignoring $\s{k}_{1,2}$ in the numerators of the propagator and using similar relations given in \eq{soft-a}. 
In the soft-photon approximation, the amplitude of diphoton+jet production in quark-nucleus collisions becomes
\bea \label{ms-di}
\langle q({\bf q}),\gamma({\bf k_1}),\gamma({\bf k_2})|q({\bf p})\rangle &\approx &ie_q^2\bar{u}({\bf q})\Big[\frac{\gamma^-
   p\cdot \epsilon(k_2)\, p\cdot \epsilon(k_1)}
{p\cdot (k_1+k_2) (p\cdot k_1)}
+ \frac{\gamma^-
   q\cdot \epsilon(k_2)\, q\cdot \epsilon(k_1)}
{(q\cdot k_2) q\cdot (k_1+k_2)}-\frac{\gamma^-
   q\cdot \epsilon(k_2)\, p\cdot \epsilon(k_1)}
{(q\cdot k_2) (p\cdot k_1)} \nonumber\\
&+&\,(k_{1}\leftrightarrow k_{2})\Big]u({\bf p}) \nonumber\\
&\times& 2\pi \delta(q^-+k^-_1+k^-_2-p^-)\int d^2{\bf z}_T\big[U({\bf z_T})-1\big] e^{i({\bf q}_T+{\bf k}_{1T}+{\bf k}_{2T}-{\bf p}_T)\cdot {\bf z}_T}.
\eea
Using the above expression for the amplitude, after  some algebra one can significantly simplify the spinor trace in \eq{tt} to obtain
\bea
&& \langle \text{tr}(S^{\dag} S)\rangle_{spin}^{\text{diphoton, soft}}  =\frac{1}{2}
\text{tr}\{\s{q}\gamma^-\s{p}\gamma^-\}
\Big{|} \frac{p\cdot \epsilon(k_2)\, p\cdot \epsilon(k_1)}
{p\cdot (k_1+k_2) (p\cdot k_1)} 
+ \frac{
   q\cdot \epsilon(k_2)\, q\cdot \epsilon(k_1)}
{(q\cdot k_2) q\cdot (k_1+k_2)}  -\frac{q\cdot \epsilon(k_2)\, p\cdot \epsilon(k_1)}{(q\cdot k_2) (p\cdot k_1)}+ (k_{1}\leftrightarrow k_{2} )\Big{|}^2,\nonumber\\
&=&\text{tr}\{\s{q}\gamma^-\s{p}\gamma^-\} (p\cdot q)^2\Bigg[\frac{1}{p\cdot(k_1+k_2)(p\cdot k_1)(q\cdot k_2)q\cdot(k_1+k_2)}+ \frac{1}{p\cdot (k_1+k_2)(p\cdot k_1)(q\cdot k_1)q\cdot (k_1+k_2)}\nonumber\\
&+&\frac{1}{(q\cdot k_2)q\cdot(k_1+k_2)p\cdot (k_1+k_2)(p\cdot k_2)}+\frac{1}{(q\cdot k_1)q\cdot (k_1+k_2)p\cdot (k_1+k_2)(p\cdot k_2)}+\frac{1}{(q\cdot k_2)(p\cdot k_1)(q\cdot k_1)(p\cdot k_2)}\Bigg],\nonumber\\
&=& 64\, p^-q^-k^{-2}_1 k^{-2}_2 q_T^4 \Bigg[\frac{k^{-}_1k^{-}_2}{\mathcal{O} M}\left(\frac{1}{k_{1T}^2 D(k_{2})}+\frac{1}{k_{2T}^2 D(k_{1})}\right) 
+\frac{1}{\mathcal{O} M}\left(\frac{k^{-2}_1}{k_{1T}^2 D(k_{1})}+\frac{k^{-2}_2}{k_{2T}^2 D(k_{2})}\right)
+ \frac{1}{k_{1T}^2 k_{2T}^2 D(k_{1}) D(k_{2})}
\Bigg],
 \label{soft-dd}\nonumber\\
\eea
where we introduced the following notation, 
\bea 
D(k_{i})&=&(k^-_i {\bf q}_T-q^- {\bf k}_{iT})^2 \,\,\,\, \text{with}  \,\,\,\, i=1,2,  \nonumber\\ 
 M&=&k^{-}_2 D(k_{1}) + k^{-}_1  D(k_{2}), \nonumber\\ 
\mathcal{O}&=&k_{1T}^2k^{-}_2 + k_{2T}^2k^{-}_1,  \label{o-d}\
\eea
with  $k_1^-$, $k_2^-$ and $q^-$ being related to the transverse momenta and pseudo-rapidities  of the produced diphoton $\eta_{\gamma_1}, \eta_{\gamma_2}$ and the jet $\eta_{h}$ via 
\begin{equation}\label{a2-0}
k_1^-=\frac{k_{1T}}{\sqrt{2}}e^{\eta_{\gamma_1}},\hspace{2cm} k_2^-=\frac{ k_{2T}}{\sqrt{2}}e^{\eta_{\gamma_2}}, \hspace{2cm} q^-=\frac{ q_{T}}{\sqrt{2}}e^{\eta_{h}}.
\end{equation} 
 In the last line of \eq{soft-dd}, we explicitly used the kinematical relations between the 4-momenta of the produced photons and the jet in the light-cone frame arising due to energy-momentum conservation (see the Appendix). 
 Substituting the above expression into \eq{m22}, the diphoton+jet  cross-section at partonic level in quark-nucleus collisions can be simplified to
\bea  \label{p-h-3}
&&\frac{d\sigma^{qA\to q(q) \gamma(k_1)\gamma(k_2) X }}{d^2{\bf k}_{1T} d\eta_{\gamma_1}d^2{\bf k}_{2T} d\eta_{\gamma_2} d^2{\bf q}_T d\eta_h }=\frac{4\alpha_{em}^2e^4_q}{(2\pi)^6\sqrt{2s}}  q^- k^{-2}_1k^{-2}_2 q^4_T \,\, \delta (x_q-\frac{k_{1T}}{\sqrt{s}} e^{\eta_{\gamma_1}}-\frac{ k_{2T}}{\sqrt{s}} e^{\eta_{\gamma_2}}-\frac{q_T}{\sqrt{s}} e^{\eta_{h}} )  \nonumber\\
&\times&\Big[\frac{k^{-}_1k^{-}_2}{\mathcal{O} M}\left(\frac{1}{k_{1T}^2 D(k_{2})}+\frac{1}{k_{2T}^2 D(k_{1})}\right) 
+\frac{1}{\mathcal{O} M}\left(\frac{k^{-2}_1}{k_{1T}^2 D(k_{1})}+\frac{k^{-2}_2}{k_{2T}^2 D(k_{2})}\right)
+ \frac{1}{k_{1T}^2 k_{2T}^2 D(k_{1}) D(k_{2})}
\Big] \nonumber\\
&\times&\int d^2{\bf r}_Td^2{\bf b}_T  e^{i({\bf q}_T+{\bf k}_{1T}+{\bf k}_{2T}).{\bf r}_T} N_F({\bf b}_T, {\bf r}_T,x_ g).\
\eea
The production at partonic level is related to 
the one in proton-nucleus collisions by convoluting \eq{p-h-3} with the quark and
antiquark distribution functions of a proton and the quark-hadron 
fragmentation function
\begin{eqnarray}\label{qh-f2}
\frac{d\sigma^{qA\to h(q^\prime) \gamma(k_1) \gamma(k_2) X}}{d^2{\bf k}_{1T} d\eta_{\gamma_1}d^2{\bf k}_{2T} d\eta_{\gamma_3}d^2{\bf q}^\prime_T d\eta_h }&=& \int^1_{z_{h}^{min}} \frac{dz_h}{z_h^2} \, 
 \int\, dx_q\, 
f (x_q,\mu_I^2)  \frac{d\sigma^{q(p)\to q(q) \gamma(k_1)\gamma(k_2) X }}{d^2{\bf k}_{1T} d\eta_{\gamma_1}d^2{\bf k}_{2T} d\eta_{\gamma_2} d^2{\bf q}_T d\eta_h }  D_{h/q}(z_h,\mu_F^2).
\end{eqnarray}
The light-cone momentum fractions $x_q, x_g, z_h$ are again related to the transverse momenta and
rapidities of the produced hadron and prompt diphoton via (see Appendix for the derivation), 
\begin{eqnarray}\label{qh-k2}
x_q&=&x_{\bar{q}}=\frac{1}{\sqrt{s}}\left( k_{1T}\, e^{\eta_{\gamma_1}}+k_{2T}\, e^{\eta_{\gamma_2}}+\frac{q^\prime_T}{z_h}\, e^{\eta_{h}}\right),\nonumber\\
x_g&=&\frac{1}{\sqrt{s}}\left(k_{1T}\, e^{-\eta_{\gamma_1}}+k_{2T}\, e^{-\eta_{\gamma_2}}+ \frac{q^\prime_T}{z_h}\, e^{-\eta_{h}}\right),\nonumber\\
z_h&=&q_T^\prime/q_T \hspace{1 cm} \text{with}~~~~~ z_{h}^{min}=\frac{q_T^\prime}{\sqrt{s}}
\left(\frac{e^{\eta_h}}
{1 - \frac{k_{1T}}{\sqrt{s}}\, e^{\eta_{\gamma_1}}-\frac{k_{2T}}{\sqrt{s}}\, e^{\eta_{\gamma_2}} }\, 
\right).\
\end{eqnarray}
One can obtain the inclusive diphoton cross-section from the semi-inclusive diphoton+jet cross-section given in \eq{ms-di} by integrating over the out-going jet momentum. To simplify the algebra, we introduce photon fragmentation parameters $z_1$ and $z_2$. The parameters $z_1$ and $z_2$ are the fraction of energy of parton carried away by produced photons with momenta $k_{1}$ and  $k_{2}$ respectively,
\be
z_1=\frac{k^{-}_1}{p^-}, \hspace{2cm}  z_2=\frac{k^{-}_2}{p^{-}-k^{-}_1}. 
 \ee
In the soft-photon limit we have $z_1 \approx \frac{k^{-}_1}{q^-}$  and $z_2\approx \frac{k^{-}_2}{q^-}$. Therefore \eq{ms-di} and \eq{p-h-3} yield 
\bea  \label{p-h-4}
&&\frac{d\sigma^{qA\to \gamma(k_1)\gamma(k_2) X }}{d^2{\bf k}_{1T} d\eta_{\gamma_1}d^2{\bf k}_{2T} d\eta_{\gamma_2} }=\frac{2\alpha_{em}^2e^4_q}{(2\pi)^6 } z_1^2 z_2^2 \int d^2{\bf q}_T\,\, q^4_T  \nonumber\\
&\times&\Bigg[\frac{k^{-}_1k^{-}_2}{\mathcal{O} \mathcal{M}}\left(\frac{1}{k_{1T}^2 \mathcal{D}(k_{2})}+\frac{1}{k_{2T}^2 \mathcal{D}(k_{1})}\right) 
+\frac{1}{\mathcal{O} \mathcal{M}}\left(\frac{k^{-2}_1}{k_{1T}^2 \mathcal{D}(k_{1})}+\frac{k^{-2}_2}{k_{2T}^2 \mathcal{D}(k_{2})}\right)
+ \frac{1}{k_{1T}^2 k_{2T}^2 \mathcal{D}(k_{1}) \mathcal{D}(k_{2})} \Bigg] \nonumber\\
&\times&\int d^2{\bf r}_Td^2{\bf b}_T  e^{i({\bf q}_T+{\bf k}_{1T}+{\bf k}_{2T}).{\bf r}_T} N_F({\bf b}_T, {\bf r}_T,x_ g),\
\eea
with $\mathcal{O}$ defined in \eq{o-d} and 
\bea
\mathcal{D}(k_{i})&=&(z_i {\bf q}_T-{\bf k}_{iT})^2 \,\,\,\, \text{with}  \,\,\,\, i=1,2, \nonumber\\ 
 \mathcal{M}&=&k^{-}_2 \mathcal{D}(k_{1}) + k^{-}_1  \mathcal{D}(k_{2}). \
\eea
 The relations between the light-cone variables $x_g, z_1, z_2$ and final state momenta for inclusive diphoton production 
 are given below  (for the derivation, see the Appendix)
\begin{eqnarray}
x_g\left(q_{T}; k_{1T}, \eta_{\gamma_{1}}; k_{2T},\eta_{\gamma_{2}}\right)&=&\frac{1}{x_q s}\Big[\frac{ k_{1T}^2}{z_1}+ \frac{k_{2T}^2}{z_2(1-z_1)}  +\frac{q_T^2}{1-z_1-z_2+z_1 z_2} \Big], \label{xg} \nonumber\\
z_1 &=& \frac{k_{1T}}{x_q\, \sqrt{s}}e^{\eta_{\gamma_1}},\nonumber\\
z_2 &=& \frac{k_{2T}}{x_q(1-z_1)\, \sqrt{s}}e^{\eta_{\gamma_2}}. \label{var}\
\end{eqnarray}

Similar to Eqs.\, (\ref{s-p},\ref{s-p-2}), one can treat the collinear divergence in the cross-section \eq{p-h-4} by introducing a hard cutoff and separating the collinear singular part into the photon fragmentation contribution.  The structure of the collinear singularity in  different terms in \eq{p-h-4} is very similar, except the last term which can be also rewritten in terms of two separated similar singular terms as long as  $\frac{{\bf k}_{1T}}{z_1} \ne \frac{{\bf k}_{2T}}{z_2}$, using the identity, 
\begin{equation} \label{int}
\frac{1}{\mathcal{D}(k_1) \mathcal{D}(k_2)} = \left( \frac{1}{\mathcal{D}(k_1)} +\frac{1}{\mathcal{D}(k_2)} \right)
\frac{1}{\mathcal{D}(k_{1})+\mathcal{D}(k_{2})}.
\end{equation} 
It is convenient to perform some variable changes in  \eq{p-h-4}.  In the terms containing the factor $1/\mathcal{D}(k_1)$ and $1/\mathcal{D}(k_2)$ we change the variable $q_T$   to ${\bf q}_T \to {\bf q}_T+\frac{{\bf k}_{1T}}{z_1}$  and  ${\bf q}_T \to {\bf q}_T+\frac{{\bf k}_{2T}}{z_2}$ respectively.  The infrared divergent part of the integral is
then extracted in the same fashion as for the single inclusive photon production in \eq{s-p-2}. After some tedious but straightforward algebra one can write the diphoton cross-section in terms of fragmentation and direct parts,
 \bea  
&&\frac{d\sigma^{qA\to \gamma(k_1)\gamma(k_2) X }}{d^2{\bf k}_{1T} d\eta_{\gamma_1}d^2{\bf k}_{2T} d\eta_{\gamma_2} }=
\frac{d\sigma^{\text{Direct}}}{d^2{\bf k}_{1T} d\eta_{\gamma_1}d^2{\bf k}_{2T} d\eta_{\gamma_2} } +\frac{d\sigma^{\text{Fragmentation}}}{d^2{\bf k}_{1T} d\eta_{\gamma_1}d^2{\bf k}_{2T} d\eta_{\gamma_2} }. \
\eea
The direct diphoton contribution is given by 
\bea 
&&\frac{d\sigma^{\text{Direct}}}{d^2{\bf k}_{1T} d\eta_{\gamma_1}d^2{\bf k}_{2T} d\eta_{\gamma_2} } =
\frac{2\alpha_{em}^2e^4_q}{(2\pi)^6 } \int_{q_T^2>\mu_F^2}d^2 {\bf q}_T\,  \frac{|{\bf q}_T+{\bf k}_{1T}/z_1|^4}{q_T^2} 
N_F\Big(|{\bf q}_T+{\bf k}_{1T}(1+1/z_1)+{\bf k}_{2T}|, x_ {g}\left(|{\bf q}_T+{\bf k}_{1T}/z_1| \right) \Big) \nonumber\\
&\times&z_2^2\Bigg[\frac{1}{ k_{1T}^2k_{2T}^2\left(q_T^2 z_1^2+z_2^2|{\bf q}_T+{\bf k}_{1T}/z_1-{\bf k}_{2T}/z_2|^2\right)}  
+
\left( \frac{k_1^-k_2^-}{\mathcal{O} k_{2T}^2}+\frac{k_1^{-2}}{\mathcal{O} k_{1T}^2}\right) \frac{1}{k_2^-q_T^2 z_1^2+k_1^-z_2^2|{\bf q}_T+{\bf k}_{1T}/z_1-{\bf k}_{2T}/z_2|^2} \Bigg]\nonumber\\
&+& (k_{1}\leftrightarrow k_{2}, z_{1}\leftrightarrow z_{2}). \label{ff-d0}\
\eea
The fragmentation contribution can be written in terms of a single and double photon fragmentation functions,  
\bea
&&\frac{d\sigma^{\text{Fragmentation}}}{d^2{\bf k}_{1T} d\eta_{\gamma_1}d^2{\bf k}_{2T} d\eta_{\gamma_2} } |_{ \frac{{\bf k}_{1T}}{z_1}\neq \frac{{\bf k}_{2T}}{z_2}} =
 \frac{\alpha_{em}e^2_q}{2(2\pi)^4 } \frac{k_{1T}^2z_2^2}{|{\bf k}_{1T}z_2-{\bf k}_{2T}z_1|^2} \Bigg[ 
\frac{1}{k_{2T}^2}  +\frac{k_{1T}^2k_2^- }{k_{2T}^2 \mathcal{O}}+\frac{k^-_1}{\mathcal{O}} \Bigg] 
\frac{1}{z_1}D_{\gamma/h}(z_1,\mu_F^2) \nonumber\\
&\times&N_F\Big(|{\bf k}_{1T}(1+1/z_1)+{\bf k}_{2T}|,x_{g}\left(k_{1T}/z_1 \right)\Big)+ (k_{1}\leftrightarrow k_{2}, z_{1}\leftrightarrow z_{2} ), \label{ff-d2}\nonumber\\
&&\frac{d\sigma^{\text{Fragmentation}}}{d^2{\bf k}_{1T} d\eta_{\gamma_1}d^2{\bf k}_{2T} d\eta_{\gamma_2} } |_{ \frac{{\bf k}_{1T}}{z_1} =\frac{{\bf k}_{2T}}{z_2}}=
 \frac{\alpha_{em}e^2_q}{2(2\pi)^4}  \Bigg[ \frac{1}{2} +\left( k_{1}^{-} k_2^-  k_{1T}^2+k_1^{-2} k_{2T}^2\right) \frac{z_2^2}{\mathcal{O}(k^-_2z_1^2+k^-_1z_2^2)}\Bigg] \frac{1}{z_1 z_2} D_{\gamma_1\gamma_2/h}(z_1,z_2, \mu_F^2)
\nonumber\\
&\times & N_F\Big( |{\bf k}_{1T}(1+1/z_1)+{\bf k}_{2T}|,x_{g}\left(k_{1T}/z_{1}\right)\Big)+(k_{1}\leftrightarrow k_{2}, z_{1}\leftrightarrow z_{2}). \label{ff-d3}\
\eea
The single photon fragmentation function $D_{\gamma/h}$ was defined in \eq{s-ff} and the diphoton fragmentation function in the soft limit in the leading-log approximation is, 
\begin{equation} \label{di-ff}
D_{\gamma_1\gamma_2/h}(z_1,z_2,\mu_F^2)=\frac{\alpha_{em} e^2_q}{\pi}\frac{1}{z_1 z_2}\left({\frac{1}{\Lambda_{QCD}^2}}-\frac{1}{\mu_F^2}\right).
\end{equation} 
In Eqs.\,(\ref{ff-d0},\ref{ff-d3}), we used a short-hand notation for the  light-cone variable 
$x_g(q_T)\equiv \ x_g\left(q_{T};\, k_{1T}, \eta_{\gamma_{1}};\, k_{2T},\eta_{\gamma_{2}}\right)$ where $x_g$ was defined in \eq{xg}. Therefore, one should bear in mind that in different terms in direct and fragmentation parts, the arguments of the dipole-target scattering amplitude $N_F\left(k_T, x_g\right)$ (the transverse momenta $k_T$ and gluon light-cone variable $x_g$) are different.

Note that as long as $\frac{{\bf k}_{1T}}{z_1}\neq \frac{{\bf k}_{2T}}{z_2}$, the two collinear singularities of the integrand in \eq{p-h-4} do not coincide, and therefore the diphoton fragmentation contribution in \eq{ff-d2} can be written in terms of two single photon fragmentation contributions. When $\frac{{\bf k}_{1T}}{z_1} \approx \frac{{\bf k}_{2T}}{z_2}$,  the  collinear singularity  in \eq{p-h-4}  is stronger than the case of the single photon production in Eqs.\,(\ref{s-p-2},\ref{s-ff}).

Both the semi-inclusive diphoton+jet and inclusive diphoton production cross-section (both direct and fragmentation part)  depend on the dipole-target amplitude and therefore in principle probe the small-x dynamics. In contrast to the dihadron production at leading-log which involves higher number of Wilson lines, the diphoton production, depends only on the dipole amplitude.  Note  that the light-cone variables $x_g$ and $x_q$ that enter the diphoton+jet and diphoton production cross sections are different, see Eqs.\,(\ref{qh-k2}, \ref{var}). Therefore, the two cross sections  in principle are sensitive to different kinematical regions of the dipole amplitude.

The production in proton-nucleus collisions is related to the above partonic cross-section via
 \begin{equation}\label{photon-f3}
\frac{d\sigma^{pA\to \gamma(k_1)\gamma(k_2) X }}{d^2{\bf k}_{1T} d\eta_{\gamma_1}d^2{\bf k}_{2T} d\eta_{\gamma_2} }
= \int^1_{x_{q}^{min}} dx_q [f_q(x_q,\mu_I^2)+f_{\bar{q}}(x_{\bar{q}},\mu_I^2)] \, 
 \frac{d\sigma^{qA\to \gamma(k_1)\gamma(k_2) X }}{d^2{\bf k}_{1T} d\eta_{\gamma_1}d^2{\bf k}_{2T} d\eta_{\gamma_2} },
\end{equation}
where
the parameter $x_q$ is the ratio of the incoming quark to the projectile nucleon energy and the lower limit of integral $x_{q}^{min}$ is defined by, 
\begin{equation}
x_q^{min} = Max\left( \frac{k_{1T} e^{\eta_{\gamma_1}}}{\sqrt{s}}, \frac{k_{2T}e^{\eta_{\gamma_2}}}{\sqrt{s}- k_{1T}e^{\eta_{\gamma_1}}} \right). 
\end{equation}


Before proceeding with numerical computation, a comment here is in order. In the soft limit, we assumed that for large $s$, the "$-$" component of the incoming projectile momentum is approximately unchanged by the interaction, and the transverse momenta of the emitted photons are small, $k_{1T}, k_{2T}\lsim q_T$ with ${\bf q}^2/s<<1$. This approximation is not appropriate for  $q_T=0$. Since the produced quark momentum is integrated over to obtain the inclusive diphoton cross section, it is essential to check that the contribution of this kinematic region is not important. Under the kinematic condition that $p^-\approx q^-$ and $q_T=0$, the trace in \eq{tt} can be  analytically calculated to give
\be\label{soft2}
 \langle \text{tr}(S^{\dag} S)\rangle_{spin}  = \frac{8\left(k_1.k_2\right)^2\left(k_1^{+}+k_2^{+}\right)^2 \Big[\left(k_1^{+}-k_2^+\right)^2p^{-2}-\left(k_1.k_2\right)^2 \Big]}{ \left(k_1^+ k_2^+\right)^2 \Big[(k_1^++k_2^+)^2 p^{-2}-\left(k_1.k_2\right)^2\Big]^2},
\ee
where for ${\bf q}_T=0$ one has  $k_1^++k_2^+ \approx x_g\sqrt{s/2}$ and $q^-=p^-\approx x_q\sqrt{s/2}$. After straightforward algebra, one can show that $z_1$ and $z_2$ dependence of the two expressions,  Eq.\, (\ref{soft2}) and Eq.\,(\ref{soft-dd}) is similar and for $z_1, z_2\to 0$ the above expression approaches zero. Moreover, for the inclusive diphoton production, the expression \eq{soft2} enters the cross section multiplied by a factor that vanishes at small $q_T$. Therefore the contribution of this kinematical region to the inclusive diphoton cross section is indeed negligible.

\section{Numerical results and discussion}
The main ingredient in the calculation of the cross-section of semi-inclusive diphoton+jet production in \eq{p-h-3},  inclusive direct and fragmentation diphoton production Eqs.\,(\ref{ff-d0},\ref{ff-d3}) is the two-dimensional Fourier transform of the universal dipole-target forward scattering amplitude $N_F$. It incorporates small-x dynamics and can be calculated by solving the non-linear JIMWLK equations \cite{jimwlk}. 
In the large $N_c$ limit, the coupled JIMWLK equations are simplified to the Balitsky-Kovchegov (BK) equation \cite{bk}, a closed-form equation for the rapidity evolution of the dipole amplitude which is presently known to next-to-leading accuracy \cite{Balitsky:2008zza,Kovner:2013ona}. The running-coupling improved BK equation (rcBK)  has the same generic formal form as the leading-log BK evolution equation:
\begin{equation}
  \frac{\partial N_{F}(r,x)}{\partial\ln(x_0/x)}=\int d^2{\vec r_1}\
  K^{{\rm run}}({\vec r},{\vec r_1},{\vec r_2})
  \left[N_{F}(r_1,x)+N_{F}(r_2,x)
-N_{F}(r,x)-N_{F}(r_1,x)\,N_{F}(r_2,x)\right],
\label{bk1}
\end{equation}
where the modified evolution kernel $K^{{\rm run}}$ using Balitsky`s
prescription \cite{bb} for the running coupling is given by,
\begin{equation}
  K^{{\rm run}}(\vec r,\vec r_1,\vec r_2)=\frac{N_c\,\alpha_s(r^2)}{2\pi^2}
  \left[\frac{1}{r_1^2}\left(\frac{\alpha_s(r_1^2)}{\alpha_s(r_2^2)}-1\right)+
    \frac{r^2}{r_1^2\,r_2^2}+\frac{1}{r_2^2}\left(\frac{\alpha_s(r_2^2)}{\alpha_s(r_1^2)}-1\right) \right],
\label{kbal}
\end{equation}
with $\vec r_2 \equiv \vec r-\vec r_1$ \cite{bb,rcbk}.
The only external input necessary for solving the rcBK non-linear equation is the initial condition for the amplitude. We take it to have the following form, motivated by McLerran-Venugopalan (MV) model \cite{mv}, 
 \begin{equation}
N(r,Y\!=\!0)=
1-\exp\left[-\frac{\left(r^2\,Q_{0s}^2\right)^{\gamma}}{4}\,
  \ln\left(\frac{1}{\Lambda_{QCD}\,r}+e\right)\right]. 
\label{mv}
\end{equation}
The infrared scale is taken as $\Lambda_{QCD}=0.241$ GeV and the onset of the small-x evolution is assumed to be at $x_0=0.01$ \cite{jav1}. The free parameters in the rcBK equation are $\gamma$ and the initial saturation scale $Q_{0s}$ (as probed by quarks), with $s=p,A$ for the proton and nuclear target, respectively. The initial saturation scale of proton $Q_{0p}^2\simeq 0.168\,\text{GeV}^2$  with the corresponding $\gamma \simeq 1.119$ was extracted from a global fit to proton structure functions in DIS in the small-x region \cite{jav1} and single inclusive hadron data in p+p collisions at RHIC and the LHC \cite{jav-d,pa-jam,pa-R}.  Note that the current HERA data alone are not enough to uniquely fix the values of $Q_{0p}$ and $\gamma$ \cite{jav1}. For the nucleus case, the initial saturation scale of a nucleus $Q_{0A}^2\approx 5 Q_{0p}^2$ should be considered as an impact-parameter averaged value and it is extracted from the minimum-bias data in deuteron-gold at RHIC and proton-lead collisions at the LHC \cite{pa-R}.

\begin{figure}       
              \includegraphics[width=8.5 cm] {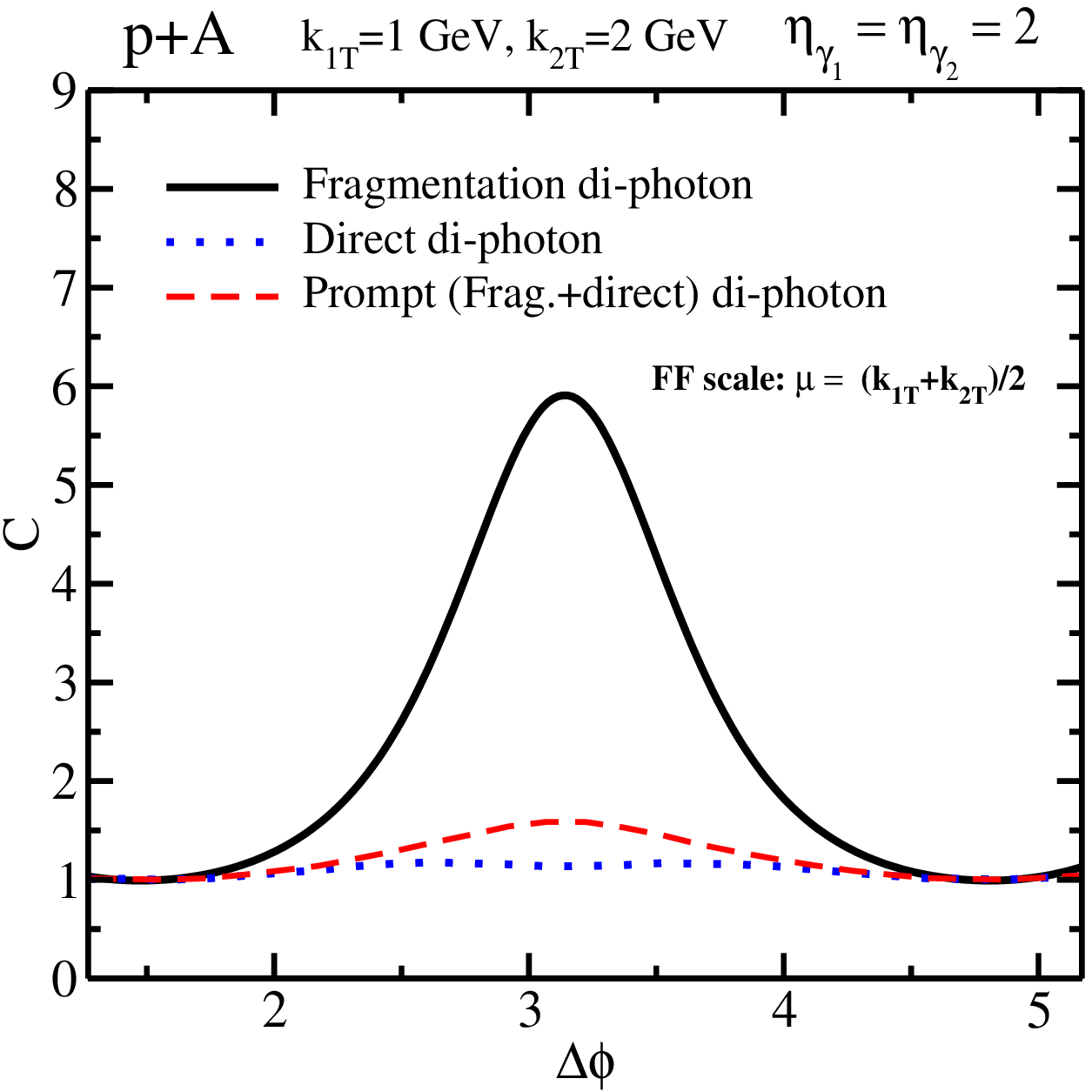}     
 \includegraphics[width=8.5 cm] {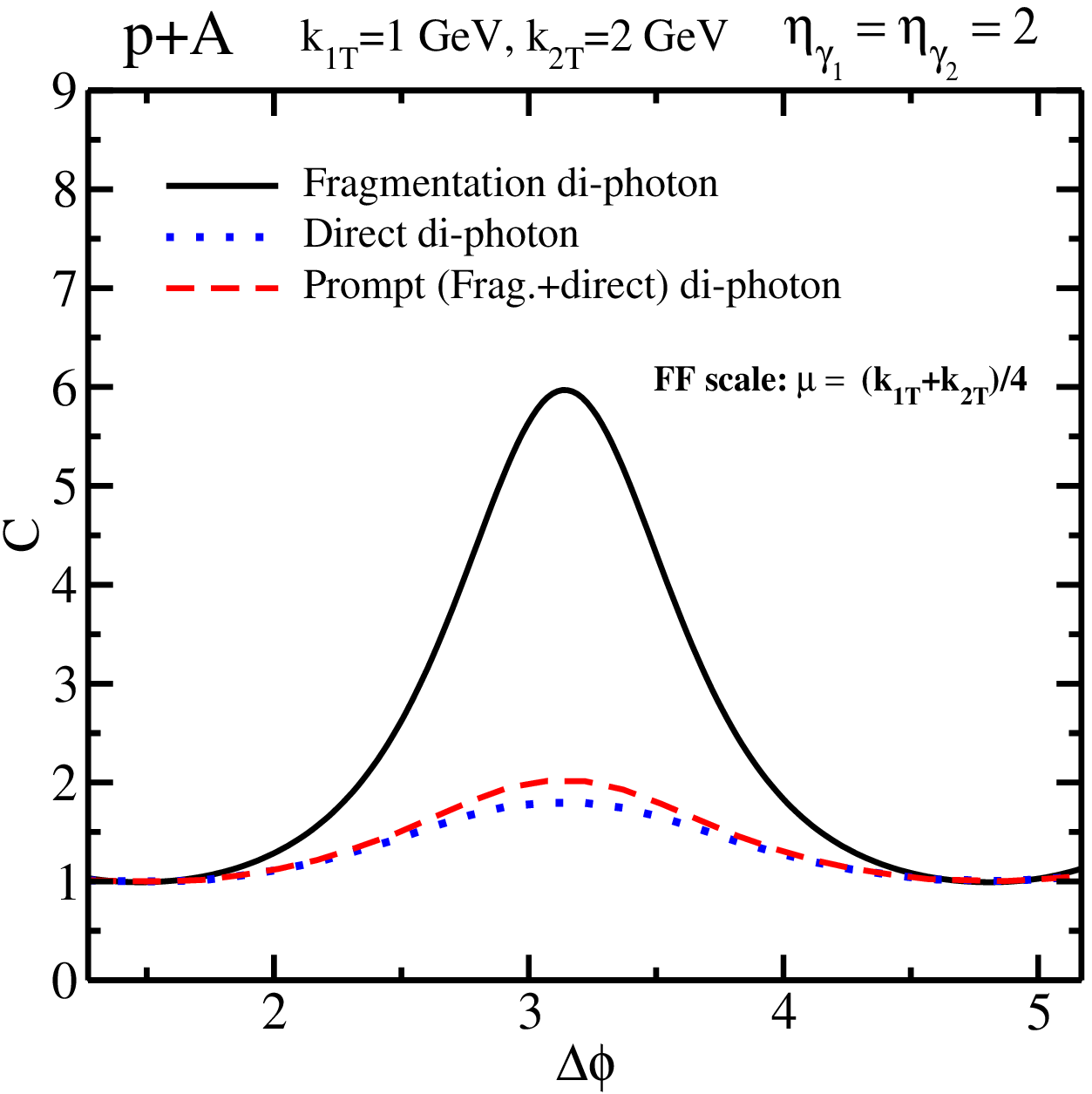}                                                                                         
\caption{Correlations of fragmentation, direct and prompt diphoton production for two different fragmentation/factorization scale  $\mu$ (left panel) and $\mu/2$ (right panel) where we defined $\mu=\mu_F=\mu_I=(k_{1T}+k_{2T})/2$. All curves are results obtained at a fixed pseudo-rapidity $\eta_{\gamma_1}=\eta_{\gamma_2}=2$  and fixed  transverse momenta $k_{1T}=1$ GeV, and $k_{2T}=2$ GeV in minimum-bias proton-lead (p+A) collisions at the LHC $\sqrt{s}=8.8$ TeV.  }
\label{dir1}
\end{figure}

Let us define the azimuthal correlation of the produced diphoton as \cite{amir-photon1,amir-photon2}, 
\begin{equation}\label{az}
C(\Delta \phi)=
\frac{d\sigma^{pA\to \gamma(k_1)\gamma(k_2) X }}{d^2{\bf k}_{1T} d\eta_{\gamma_1}d^2{\bf k}_{2T} d\eta_{\gamma_2} }[\Delta \phi]/\frac{d\sigma^{pA\to \gamma(k_1)\gamma(k_2) X }}{d^2{\bf k}_{1T} d\eta_{\gamma_1}d^2{\bf k}_{2T} d\eta_{\gamma_2} } [\Delta \phi= \Delta \phi_c],
\end{equation}
where $\Delta \phi$  is the azimuthal angle between the two produced photons in the plane transverse to the collision axis. The azimuthal correlation $C$ is proportional to the probability of inclusive diphoton pair production in a given kinematics and angle  $\Delta \phi$ between the photons in the pair, normalized to the 
fixed reference angle $\Delta \phi_c$.  Since we are mostly interested to study correlations at  $\Delta \phi\approx \pi$, we fix the reference angle $\Delta \phi_c=\pi/2$ throughout this paper. 

 One can equally well  take the normalization in \eq{az}  as the differential cross-section integrated over the angle  $\Delta \phi$. We expect that some of the theoretical uncertainties, such as sensitivity to possible $K$-factors which effectively incorporates the missing higher order corrections, drop out in the correlation defined in \eq{az}. One should bear in mind that the correlation defined in \eq{az} may be more challenging to measure compared to the so-called coincidence probability \cite{di-e,ph-ex,amir-photon2} due to possible underlying event dependence, however since  it is free from  extra integrals over transverse momenta, it should exhibit  the underlying dynamics of the correlation in a cleaner way. In a sense, the correlation defined in  \eq{az}  is a snapshot of the integrand in the coincidence probability. 

\begin{figure}[t]                                       
                                  \includegraphics[width=8.5 cm] {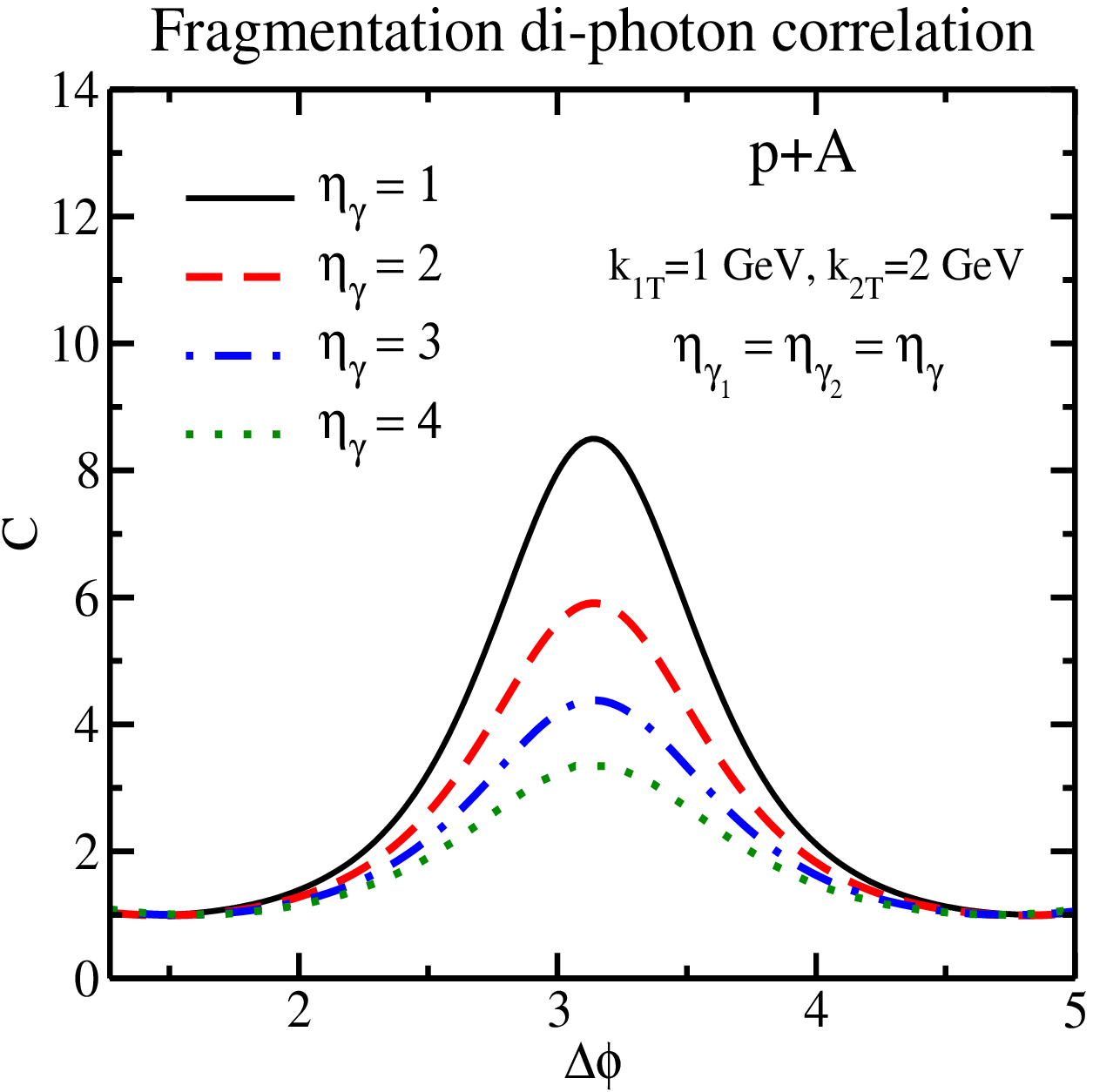} 
                                  \includegraphics[width=8.5 cm] {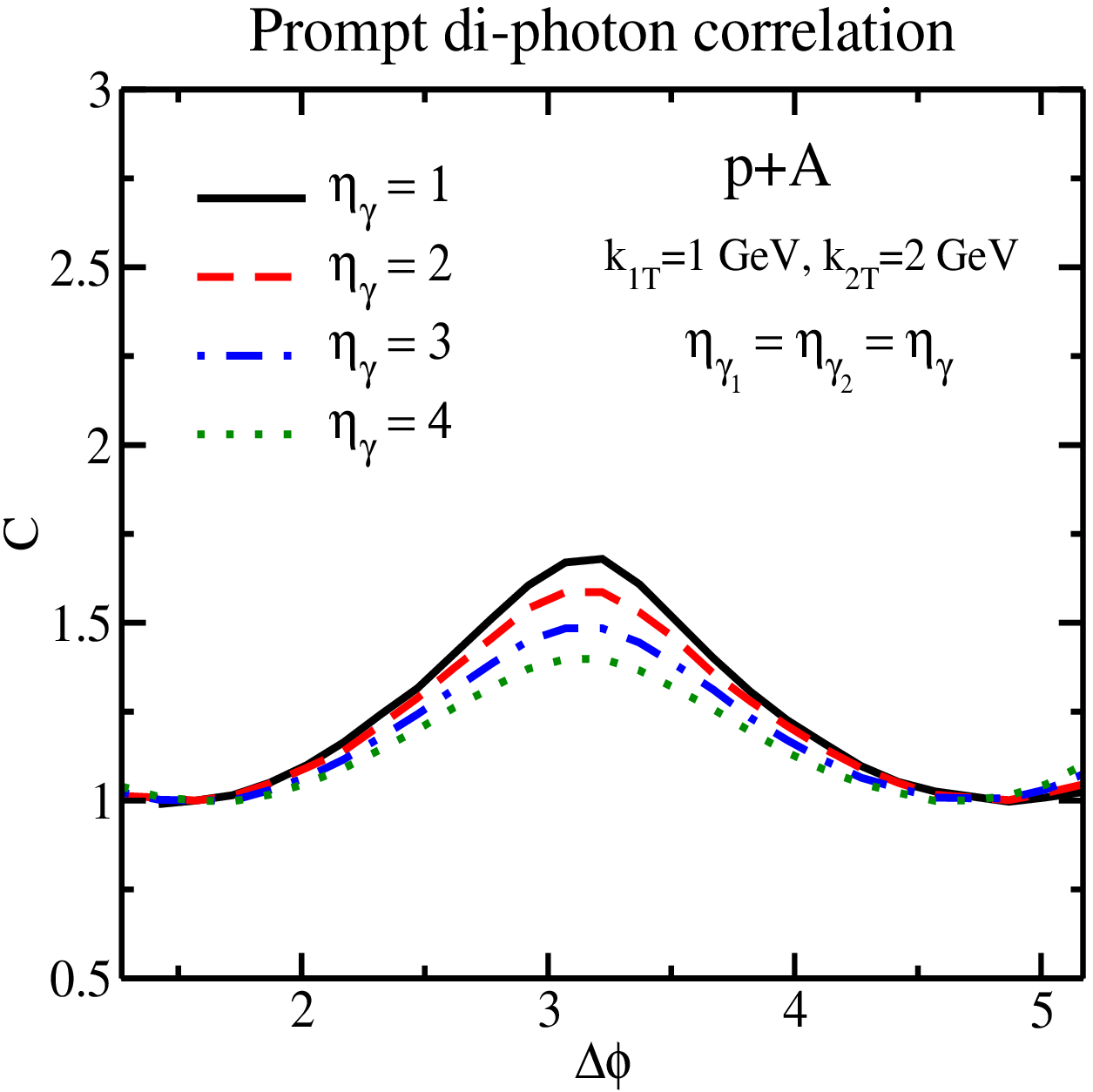}                     
\caption{Fragmentation (left panel) and prompt (right panel) diphoton correlations at different pseudo-rapidities of the produced diphoton $\eta=\eta_{\gamma_1}=\eta_{\gamma_2}$ in minimum-bias proton-lead collisions. 
In both panels,  the curves are the results obtained from the rcBK evolution equation with transverse momenta of diphoton, fixed at $k_{1T}=1$ GeV, and $k_{2T}=2$ GeV at LHC energy $\sqrt{s}=8.8$ TeV. In both panels, the factorization and fragmentation scales are taken to be equal $\mu_F=\mu_I=(k_{1T}+k_{2T})/2$.    }
\label{frag1}
\end{figure}

\begin{figure}[t]                                       
                                  \includegraphics[width=8.5 cm] {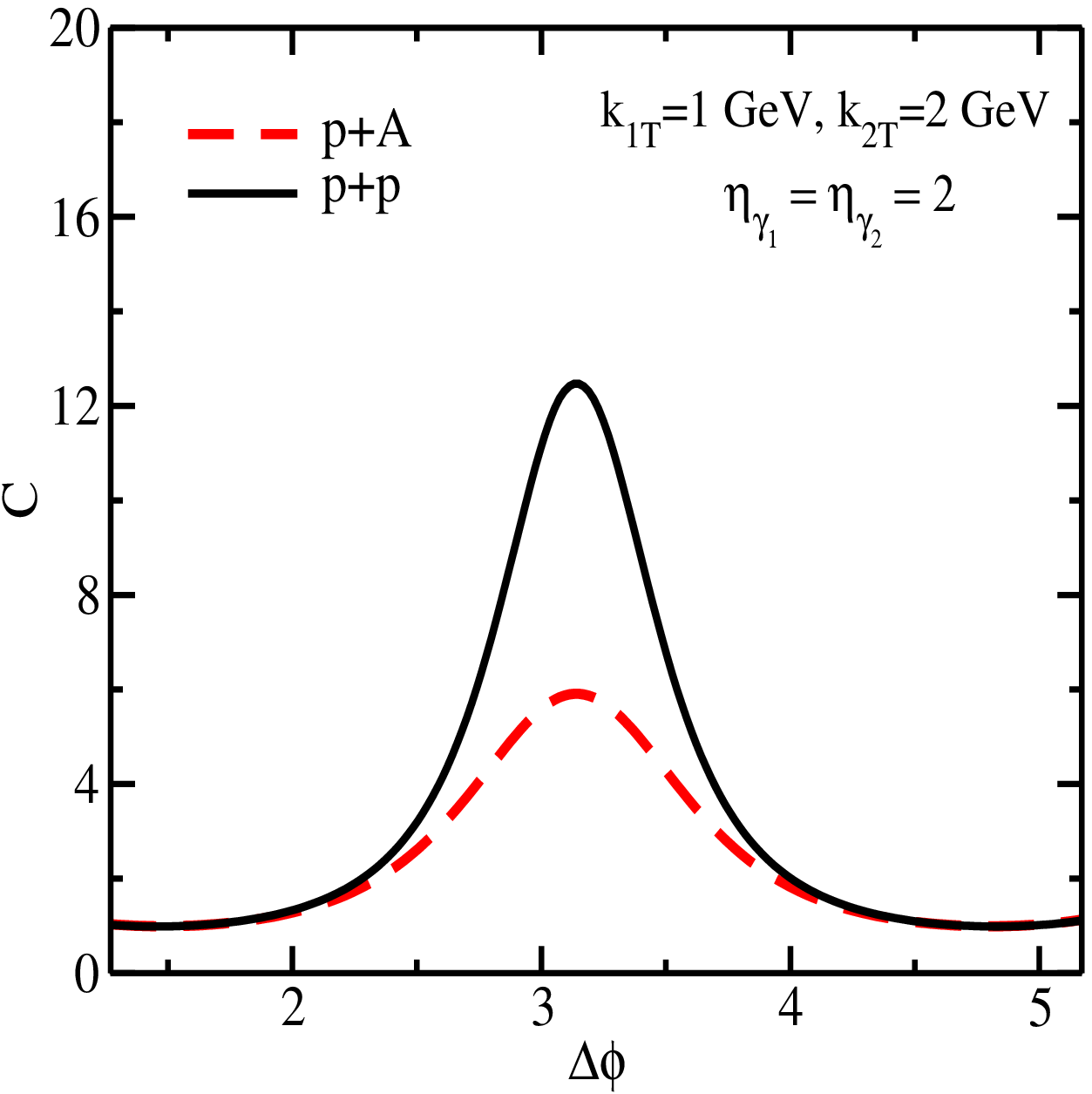}                     
\caption{ Comparison of fragmentation diphoton correlations in minimum-bias proton-proton and proton-lead collisions at a fixed pseudo-rapidity $\eta_{\gamma_1}=\eta_{\gamma_2}=2$ with $k_{1T}=1$ GeV, and $k_{2T}=2$ GeV at the LHC energy $\sqrt{s}=8.8$ TeV.   }
\label{frag1-pa}
\end{figure}

Note that the hard scale $\mu_I$ in the parton distribution in \eq{photon-f3} can be in principle different from the photon fragmentation scale $\mu_F$ introduced in Eqs.\,(\ref{ff-d0}-\ref{ff-d3}). Following the conventional pQCD approach, we take the hard scale $\mu_I$ to be  equal to the fragmentation scale $\mu_F$, namely $\mu=\mu_F=\mu_I$. 
We will later quantify uncertainties associated with the freedom to choose a different  scale $\mu$. For the patron distributions, we will use the NLO MSTW PDFs \cite{mstw}.  
For numerical computation, we focus here at low transverse momenta of the produced photon pair at the LHC at forward rapidities, consistent with the soft approximation employed for obtaing the cross-section.  Note that this kinematics is mostly relevant for probing saturation effects.

In the standard perturbative calculations the leading contribution to the diphoton production comes from the annihilation diagram. 
This process produces back to back photon pairs, and thus leads to strong peak in the correlation at $\Delta\phi=\pi$.
This contribution is absent in the CGC approach, since the dense target wave function is  dominated by gluons. In the CGC framework the annihilation contribution only appears in the next order in $\alpha_s$. Nevertheless, we expect the fragmentation diphoton to have a non-negligible correlation peaked at $\Delta \phi=\pi$. The reason is that the photon-quark Fock component of the incoming quark has zero transverse momentum in the collinear factorization approach. Therefore if momentum transfer from the target is small enough, the photon collinear to the outgoing quark, and the photon emerging from the initial photon-quark state will have opposite transverse momenta, leading to back-to-back correlation.

The direct diphoton part on the other hand, is restricted to kinematics where the transverse momentum of the outgoing quark jet is relatively large $q_T>\mu_F$. One therefore does not expect significant back-to-back correlation in the direct photon contribution and also expects that the correlations in the direct part should be more sensitive to the fragmentation scale. Reducing the scale $\mu$ should enhance the back-to-back correlations for the direct diphoton\footnote{Note that this is one of the main differences between diphoton and dihadron correlations. In the later case, the dihadron can be produced from splitting a single gluon and the back-to-back production is in principle kinematically allowed.}.

Indeed, our numerical results shown in \fig{dir1} follow these expectations. Note that because of the convolution with fragmentation and parton distribution functions the  partonic level correlation gets somewhat  smeared out.  In \fig{dir1}, we show that the correlations at $\Delta \phi=\pi$ in the direct diphoton contribution are indeed much smaller than in the fragmentation one.  In \fig{dir1},  we also show the effects of different fragmentation/factorization scale. We present the correlations in different components of diphoton production calculated with two different scales $\mu$ (left panel) and  $\mu/2$ (right panel).

 The fragmentation contribution is less sensitive to the choice of fragmentation/factorization scale while the correlations in the direct diphoton  are affected by this uncertainty. This is easy to understand, since the fragmentation/factorization scale that appears in the FF and the PDF of fragmentation cross-section in \eq{ff-d3} mainly cancels between the numerator and the denominator  in the correlation defined in \eq{az}. This does not happen in the  direct part since the fragmentation scale appears as the lower limit of integral in the cross-section \eq{ff-d0}.

In \fig{dir1}, we also compare the correlations of direct, fragmentation and prompt (fragmentation+direct) diphoton production at fixed kinematics at the LHC in proton-nucleus collisions for two different fragmentation/factorization scales. One notes that the back-to-back correlation is larger in the fragmentation part, while the total  (and near-side) direct diphoton cross-section is larger than the fragmentation one. 
 As a result, the total prompt diphoton signal (the sum of direct and fragmentation parts) exhibits a reduced back-to-back correlation defined via \eq{az}.  However,  with isolation cut technique \cite{di-cut} (see also Ref.\,\cite{pho-cut}), one can in principle isolate the fragmentation diphoton contribution and study this correlation separately\footnote{This would be the opposite of what one may do in order to study the direct diphoton (or photon) production by imposing isolation cut to discard the fragmentation contribution. An incorporation of the isolation cut criterion in our framework is beyond the scope of the current paper.}. Note also that by default, the back-to-back correlation is significantly larger in the two single-photon fragmentation than double-photon fragmentation part, see \eq{ff-d3}.

In \fig{frag1}, we show the rapidity dependence of the fragmentation and prompt (direct+fragmentation) diphoton correlation $C$ defined in \eq{az} at fixed transverse momenta of the produced diphoton in minimum-bias proton-nucleus collisions at the LHC energy $\sqrt{s}=8.8$ TeV. The back-to-back correlations are systematically suppressed at forward rapidities (larger $\eta_{\gamma_1}$ and  $\eta_{\gamma_2}$) in both fragemtation and prompt (and direct) diphoton production.

Given that the correlation in the direct diphoton production is small (see Figs.\,\ref{dir1},\ref{frag1}), in the following we only show the correlation calculated from the fragmentation diphoton part. The general features of the (de)-correlations discussed below, persist for the prompt diphoton production, albeit the magnitude of the correlation is uniformly smaller. We also fix the factorization/fragmentation scale to $\mu = (k_{1T}+k_{2T})/2$, as the variation of the scale does not greatly affect the correlations in the fragmentation part, see \fig{dir1}. 

In \fig{frag1-pa}, we compare the diphoton correlations in minimum-bias p+p and p+A collisions at forward rapidity $\eta_{\gamma_1}=\eta_{\gamma_2}=2$ for fixed transverse momenta of the pair at $k_{1T}=1$ GeV, and $k_{2T}=2$ GeV.  The back-to-back correlations in p+A collisions are clearly suppressed compared to p+p collisions. 

\begin{figure}[t]        
              \includegraphics[width=8.5 cm] {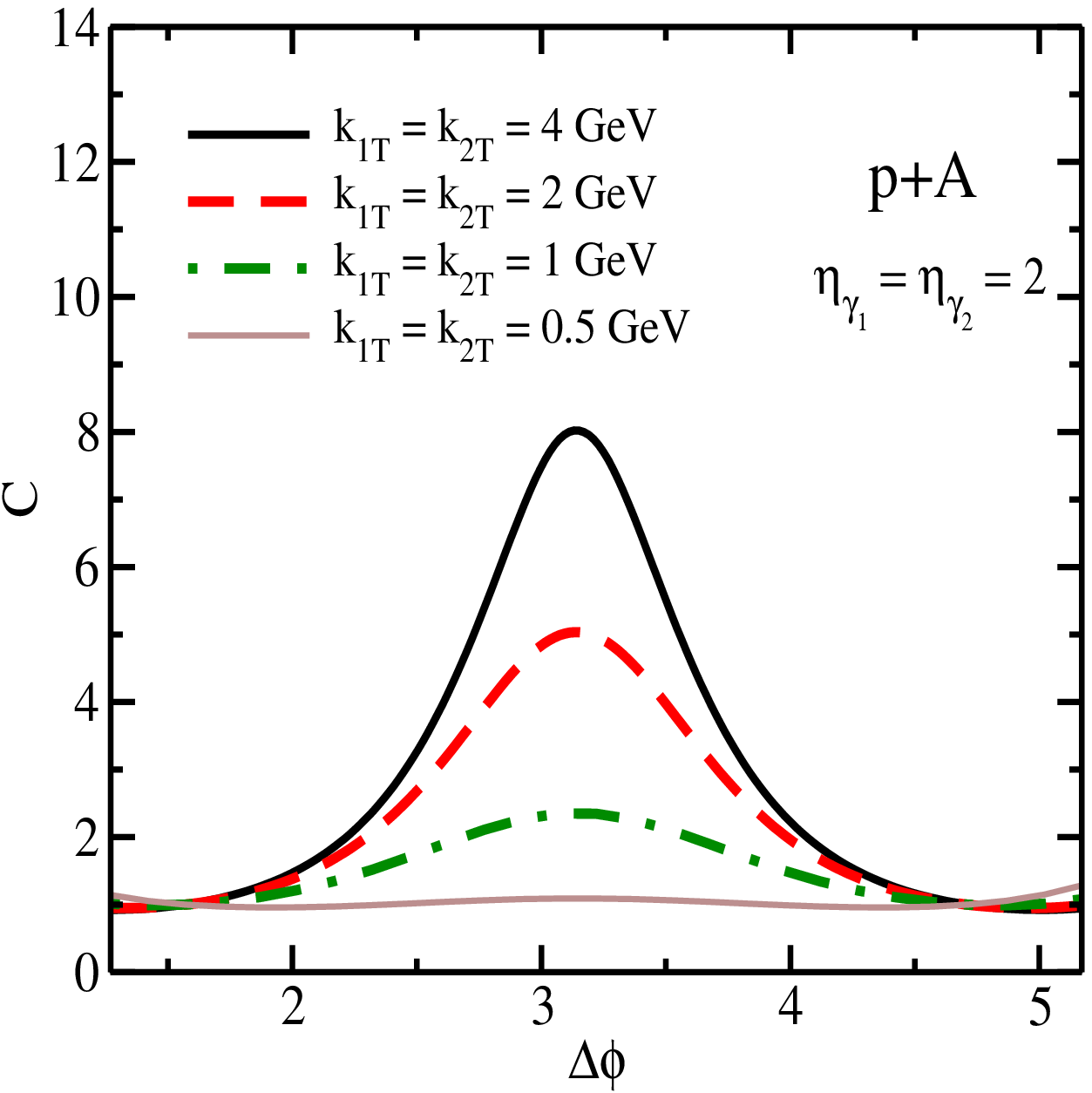}                                 
                                \includegraphics[width=8.5 cm] {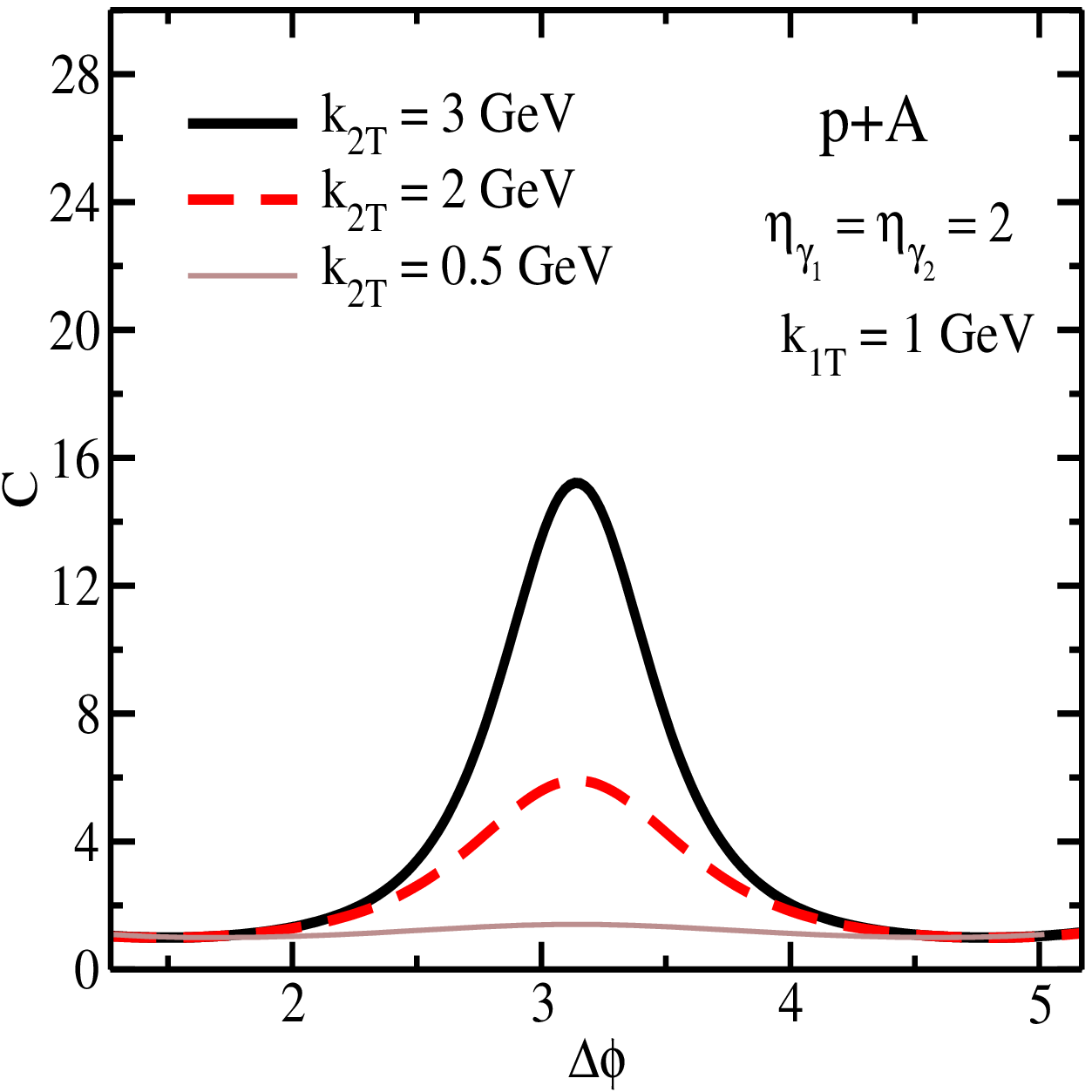}                                
\caption{Diphoton (fragmentation) correlations at different transverse momenta of the produced photon pair $k_{1T}$ and $k_{2T}$ in minimum-bias proton-lead collisions. 
In both panels, the curves are the results obtained at a fixed pseudo-rapidity $\eta_{\gamma_1}=\eta_{\gamma_2}=2$ in proton-lead collisions at the LHC $\sqrt{s}=8.8$ TeV.
     }
\label{frag2}
\end{figure}

 In \fig{frag2}, right panel, we show the effect of variation of transverse momenta of the produced photons at fixed rapidity in p+A collisions at the LHC. Lowering the transverse momenta leads to suppression of the away-side diphoton correlation. 

Note that all the features seen in Figs.\,\ref{frag1}-\ref{frag2} can be understood in the saturation picture. By increasing density or rapidity/energy or decreasing transverse momenta, the typical $x_g$ which enters in the dipole-target scattering amplitude, becomes smaller and consequently  the typical saturation scale of the system becomes larger. In this case, the intrinsic back-to-back correlation is smeared due to momentum exchange with the target at the saturation scale. This increased decorrelation with increasing saturation scale appears to be a universal feature of diphoton production, irrespective of the mechanism by which the saturation scale is increased.

\section{Conclusion}

In this paper, we investigated  semi-inclusive diphoton+jet and inclusive diphoton production at leading log approximation in high-energy proton-nucleus collisions using the Color Glass Condensate formalism. We obtained the inclusive prompt diphoton cross-section in terms of  fragmentation and direct diphoton contributions while  the fragmentation part is given in terms of single-photon  and  double-photon fragmentation functions. 

We have also studied the diphoton azimuthal angular correlations in p+p and p+A collisions at the LHC kinematics.   It is generally seen that at low transverse momenta of the produced diphoton, back-to-back correlations of fragmentation, direct and prompt diphoton production are all sensitive to saturation physics, although this sensitivity is significantly stronger in two single-photon fragmentation parts.  It was shown that the away-side peak in diphoton angular correlation is reduced by lowering the diphoton transverse momenta. At a fixed transverse momenta, the suppression of the away-side correlations gets stronger as one goes to larger rapidities (or higher energies) or a denser system.  The main features of away-side decorrelation of diphoton production seem to be universally similar to that in dihadron \cite{dihadron} and photon-hadron \cite{amir-photon1,amir-photon2} productions in high-energy p+A collisions. In all cases, the away-side correlations of the produced di-particle get suppressed in the presence of a large saturation scale irrespective of mechanism by which the particles are produced and the saturation scale is enhanced. We recall that  diphoton production is a theoretically cleaner probe of initial-state effects and small-x dynamics compared to dihadron production, mainly due to the fact that the diphoton production is free from hadronization corrections which  theoretically are not too well understood. Moreover, since the virtual photons do not interact with the gluons inside target, final-state effects are absent in the diphoton production.

\appendix
\section{}

The purpose of this appendix is to define the kinematics and derive
the needed relations between various light-cone energy fractions 
which appear in the production cross sections used. This is slightly 
different from the standard relations used in production cross sections
based on collinear factorization theorems of pQCD. We first consider
scattering of a quark on the target where a photon and a 
quark are produced, depicted in Fig.~1, 
\be
q(p)+A (p_A)\to \gamma(k_1) +\gamma(k_2) + \text{jet}(q) + X,
\ee
where $A$ is a label for the multi-gluon state, described by a classical field representing 
a proton or nucleus target. In the standard pQCD (leading twist)
kinematics, only one parton from the target interacts. This is not the case here since
the target is described by a classical gluon field representing a multi-gluon state 
with intrinsic momentum rather than an individual gluon with a well defined energy 
fraction $x_g$ and zero transverse momentum. Nevertheless, since most of the gluons in 
the target wave function have momentum of order $Q_s$, one can think of the state describing 
the target as being labeled by a (four) momentum $p_A$. In this sense, the 
gluons in the target collectively carry fraction $x_g$ of the target energy and have 
intrinsic transverse momentum denoted by ${\bf p}_{A}$. This also means that there is no integration
over $x_g$ in our case unlike the collinearly factorized cross sections in pQCD (this basically
corresponds to setting $x_g$ equal to the lower limit of $x_g$ integration in pQCD cross sections).
We thus have 
\begin{eqnarray}\label{a1}
p^\mu&=&\left(p^-=x_q\sqrt{s/2},~p^+=0,~{\bf p}_T =0 \right),\nonumber\\
P^\mu&=&\left(P^-=\sqrt{s/2},~P^+=0,~{\bf P}_T=0 \right),\nonumber\\
p_A^\mu&=&\left(p^{-}_A=0,~p_A^{+}=x_g\sqrt{s/2},~{\bf p}_{AT} \right),\nonumber\\
P_A^\mu&=&\left(P^{-}_A=0,~P_A^{+}=\sqrt{s/2},~{\bf P}_{AT}=0 \right),\nonumber\\
q^\mu&=&\left(q^- ,~q^+=q_T^2/2q^-,~{\bf q}_T \right),\nonumber\\
k_1^\mu&=& \left(k_1^-=z_1p^-=z_1x_q\sqrt{s/2},~k_1^+=k_{1T}^2/2k^-_1,~{\bf k}_{1T}\right), \nonumber\\
k_2^\mu&=& \left(k_2^-=z_2(p^--k_1^-)=x_q z_2(1-z_1)\sqrt{s/2},~k_2^+=k_{2T}^2/2k^-_2,~{\bf k}_{2T}\right), \
\end{eqnarray}
where $P^\mu , P_A^\mu, q^\mu$ are the momenta of the incoming projectile, target
and the produced jet respectively. (Pseudo)-rapidities of the produced diphoton is related to their energies via 
\begin{equation}\label{a2}
k_1^-=\frac{k_{1T}}{\sqrt{2}}e^{\eta_{\gamma_1}},\hspace{2cm} k_2^-=\frac{ k_{2T}}{\sqrt{2}}e^{\eta_{\gamma_2}}, \hspace{2cm} q^-=\frac{ q_{T}}{\sqrt{2}}e^{\eta_{h}}.
\end{equation} 
Imposing energy-momentum conservation at the partonic level via $\delta^4 (p + p_A - q -k_1-k_2)$ 
and using \eq{a1} leads to
\begin{eqnarray} \label{con}
p^-&=&k^-_1 + k^-_2+ q^-,\label{a31} \\
p^{+}_A&=&k^+_1 +k^+_2+ q^+,\label{a32}\\
{\bf p}_{AT} &=& {\bf k}_{1T} + {\bf k}_{2T}+{\bf q}_T.
\end{eqnarray}
Plugging the definitions given in \eq{a2} into the above relations and using  \eq{a1} (and the on mass-shell condition), one can immediately obtain the energy fractions $x_q,x_g$ in the case of the diphoton+jet production:
\begin{eqnarray} \label{con2}
x_q&=&x_{\bar{q}}=\frac{1}{\sqrt{s}}\left( k_{1T}\, e^{\eta_{\gamma_1}}+k_{2T}\, e^{\eta_{\gamma_2}}+q_T\, e^{\eta_{h}}\right),\nonumber\\
x_g&=&\frac{1}{\sqrt{s}}\left(k_{1T}\, e^{-\eta_{\gamma_1}}+k_{2T}\, e^{-\eta_{\gamma_2}}+ q_T\, e^{-\eta_{h}}\right),\
\end{eqnarray}
where the first and second equation was directly derived from \eq{a31} and \eq{a32}, respectively.  Note that light-cone momentum fraction $x_g$ appears in
the dipole forward scattering amplitude $N_F (b_t, r_t, x_g)$ whereas $x_q$ is the fraction of the projectile
proton carried by the incident quark, see \eq{a1}. One can relate the transverse momentum of the fragmented hadron  $q^\prime_T$ to  the out-going quark $q_T$ via $z_h=q^\prime_T/q_T$. Now using \eq{a31}, the minimum value of $z_h$ is obtained for the maximum value of $x_q=1$. Therefore, we obtain
\begin{equation}
z_h^{min}=\frac{q^{\prime -}}{\sqrt{s/2}-k_1^{-}-k_2^{-}}.
\end{equation}
For obtaining the diphoton production, one integrates over the out-going quark transverse momentum  and rapidity of diphoton+jet cross-section. The integral over rapidity of out-going jet or $q^-$ can be done analytically. Therefore, some extra care is in order here. Let us first introduce the parameter $z_1$ and $z_2$ as the fraction of energy of parton carried away by two produced photons defined by,
\bea
z_1 &\equiv& \frac{k^-_1}{p^-} = \frac{k_{1T}}{x_q\, \sqrt{s}}e^{\eta_{\gamma_1}}, \nonumber\\
z_2 &\equiv& \frac{k^-_2}{p^--k^-_1} = \frac{k_{2T}}{x_q(1-z_1)\, \sqrt{s}}e^{\eta_{\gamma_2}}.\ \label{zz}
\eea
Plugging the above relations into Eqs.\, (\ref{a31},\ref{a32}) and using \eq{a1} one can derive the following expressions for the energy fractions $x_q,x_g$, 
\begin{eqnarray}\label{a4}
x_q&=&x_{\bar{q}}=\frac{q^-}{\sqrt{s/2}\left(1-z_1-z_2+z_2 z_1 \right)},\\
x_g&=&\frac{1}{x_q s}\Big[\frac{ k_{1T}^2}{z_1}+ \frac{k_{2T}^2}{z_2(1-z_1)}  +\frac{q_T^2}{1-z_1-z_2+z_1 z_2} \Big].\label{a7}\
\end{eqnarray}
To derive an expression for 
the lower limit of $z_1$ and $z_2$ (in integration),  we note that $0 \le \, x_q \, \le \, 1$. Using 
the relations in \eq{zz}, we obtain
\bea \label{min-z}
z_{1}^{min} &=& \frac{k_{1T} e^{\eta_{\gamma_1}}}{\sqrt{s}}, \nonumber\\
z_{2}^{min} &=& \frac{k_{2T}e^{\eta_{\gamma_2}}}{\sqrt{s}(1-z_{1}^{min})}= \frac{k_{2T}e^{\eta_{\gamma_2}}}{\sqrt{s}- k_{1T}e^{\eta_{\gamma_1}}}. \
\eea
Equally, we can immediately obtain the lowest value of $x_q$ denoted by $x_q^{min}$ from Eqs.\,(\ref{zz},\ref{min-z}) by imposing the condition that $0<z_1, z_2<1$,
\begin{equation}
x_q^{min} =Max\left( \frac{k_{1T} e^{\eta_{\gamma_1}}}{\sqrt{s}}, \frac{k_{2T}e^{\eta_{\gamma_2}}}{\sqrt{s}- k_{1T}e^{\eta_{\gamma_1}}}  \right). 
\end{equation}

\begin{acknowledgments}
The authors would like to thank Jamal Jalilian-Marian for fruitful discussions at early stage of this work. A. K. thanks Physics Department of Universidad T\'ecnica Federico Santa Mar\'\i a for hospitality. The work of A.H.R. is supported in part by Fondecyt grant 1110781. The work of A.K. is supported by the DOE grant DE-FG02-13ER41989.
\end{acknowledgments}


\end{document}